\documentclass[a4paper,11pt]{article}
\pdfoutput=1 
\usepackage{jheppub}
\usepackage{amsmath}
\usepackage{amssymb}
\usepackage{color}
\usepackage{xcolor}
\usepackage{graphicx}
\usepackage{hyperref}
\usepackage{textcomp}
\usepackage{xspace,slashed}
\usepackage[utf8]{inputenc}
\usepackage{subcaption}
\usepackage{multirow}
\usepackage{nccmath} 
\usepackage{slashed}
\usepackage[bb=boondox]{mathalfa} 
\usepackage{orcidlink}
\usepackage{longtable}
\usepackage{titlesec}
\usepackage{slashed}

\makeatletter
\renewcommand\@fpheader{}
\renewcommand\@journal{}
\makeatother

\newcommand{\ep}{\epsilon}

\newcommand{\C}{\mathcal{G}}

\title{All-loop four-quark Bethe-Salpeter kernel}

\author[a]{Piotr Bargie\l{}a~\orcidlink{0000-0002-3646-5892}}
\emailAdd{pbargiel@ed.ac.uk}
\affiliation[a]{Higgs Centre for Theoretical Physics, School of Physics and Astronomy, The University of Edinburgh, Edinburgh EH9 3FD, Scotland, UK}

\preprint{}

\abstract{
We analytically calculate the all-loop bare perturbative part of the four-quark Bethe-Salpeter kernel using modern scattering amplitude methods.
We work to subleading order in the large number of quark flavors approximation of massless Quantum Chromodynamics, which simultaneously makes an all-loop calculation feasible, is systematically improvable, and preserves asymptotic freedom.
It also allows for avoiding the ambiguity of choosing a truncation scheme in Dyson-Schwinger equations.
We exploit state-of-the-art methods in Integration-By-Parts reduction of Lorentz scalar Feynman integrals into a minimal Master Integral basis, and direct integration into Generalized Polylogarithms.
As a byproduct of our calculation, we also provide the result for the gluon and quark propagators.
We discuss a path towards nonperturbative formulation and potential future phenomenological applications.
}

\begin{document}

\maketitle 

\allowdisplaybreaks

\newpage
\section{Introduction}
\label{sec:intro}

Understanding confinement of quarks and gluons into hadrons is one of the most important open problems in Theoretical Particle Physics.
Formally, these particles are governed by Quantum Chromodynamics (QCD) which describes the strong force between the partons.
This non-abelian Yang-Mills Quantum Field Theory (QFT) in $d_0=4$ dimensions with SU($N_c$) gauge group has $N_c=3$ colors and $N_f=6$ flavors of massive Dirac quark fields in the fundamental representation of the gauge group.
In contrast to abelian QFTs, QCD exhibits asymptotic freedom at high ultraviolet (UV) energies and is strongly coupled at low infrared (IR) energies.
As such, it admits perturbative expansion in the small coupling at high energies and becomes strongly interacting at low energies.
This property has been proven in 1973~\cite{Gross:1973id,Politzer:1973fx} and the authors Gross, Wilczek and Politzer have been awarded The Nobel Prize in Physics 2004. 
However, strong coupling in the IR is only indicative but not sufficient for confinement.
For a comprehensive recent review on the status of QCD 50 years after its formulation see Refs~\cite{Gross:2022hyw,Shuryak:2026pqt}.

Although confinement of colorful UV partonic degrees of freedom into colorless IR hadrons is experimentally well-established, proving it analytically remains a challenge due to its nonperturbative nature.
Indeed, in the theoretical description, one needs to take into account all orders of perturbative corrections stemming from multi-loop Feynman diagrams as well as purely nonperturbative effects. 
Therefore, the nonperturbative properties of hadrons cannot be predicted from first principles using only fixed-order perturbative QCD.
Among such important physical nonperturbative quantities are e.g. the spectrum of masses of mesons and baryons, including the lowest mass gap states like pion and proton, Regge trajectories in spin on which these masses lie, hadronic decay constants appearing in Chiral Perturbation Theory ($\chi$PT) such as pion decay constant $f_\pi$ appearing in the pion Lagrangian, and Parton Distribution Functions (PDFs) which describe the probability density of finding a given parton in the corresponding hadron at fixed momentum fraction and energy scale.

Famously, one of the seven Millennium Prize problems selected by the Clay Mathematics Institute in 2000 is to prove the existence of the mass gap in the pure Yang-Mills theory, i.e. to show that the mass of the lowest-energy hadron in a QFT with non-abelian gauge group and without matter quark fields is nonzero~\cite{Jaffe:2000}.
One of the first historical attempts to analytically calculate the mesonic mass spectrum of QCD was that of 't~Hooft in 1974~\cite{tHooft:1974pnl}.
By approximating QCD to leading order in large number of colors $N_c$~\cite{tHooft:1973alw} in $d_0=2$ dimensions, he solved the Dyson-Schwinger equation (DSE) for the nonperturbative quark propagator and used it in the four-quark kernel of the Bethe-Salpeter equation (BSE) for the meson bound state.
Extending his method to QCD in $d_0=4$ dimensions is the main motivation for this work.

Throughout the last decades, there has been a lot of progress and results in numerically solving the DSEs and BSE for QCD in $d_0=4$ dimensions, see e.g. Refs~\cite{Sterman:1973jda,Lepage:1978hz,vonSmekal:1997ern,Maris:2003vk,Aguilar:2004sw,Fischer:2006ub,vanBaalen:2009hu,Swanson:2010pw,Fu:2025hcm,Liu:2025ldh}.
However, all of these solutions suffer from an ambiguous choice of the truncation scheme of the infinite system of coupled DSEs.
The fact that there is no clear parameter in which the uncertainty resulting from a chosen truncation scheme can be systematically decreased poses a great theoretical challenge.
An alternative method based on Lattice QCD (LQCD) offers such explicit parameters, e.g. the lattice spacing and volume, which control the precision of the numerical prediction.
For this reason, there has been a lot of hadronic predictions provided within LQCD, see an example review in Refs~\cite{Hansen:2019nir,FlavourLatticeAveragingGroupFLAG:2024oxs}.

In this work, we propose an alternative path towards providing nonperturbative QCD predictions based on first principles.
We outline a construction of the gluon propagator, quark propagator, and Bethe-Salpeter (BS) kernel that relies on summing all the contributing higher-loop perturbative corrections and using the transseries resurgence techniques to retrieve the corresponding nonperturbative part, instead of solving the DSEs.
Explicitly, we detail the calculation of the all-loop perturbative part in this manuscript and leave the nonperturbative investigation for the future.
We advocate that using modern methods developed over the last few decades in the field of perturbative scattering amplitudes opens a new systematic avenue in this direction.

The scattering amplitude describes the probability amplitude for a scattering process of particles in a given QFT.
Since bare multi-loop perturbative QCD amplitudes are usually divergent, they require introducing a regularization scheme.
In this work, we will use dimensional regularization (dimReg) in $d=d_0-2\ep$ dimensions around $d_0=4$ and with $\ep \to 0$.
We follow the standard workflow of powerful modern methods in amplitude computations.
It includes the decomposition of all the corresponding Feynman diagrams into a basis of Lorentz tensors, an Integration-By-Parts (IBP)~\cite{Tkachov:1981wb,Chetyrkin:1981qh} reduction of the resulting Lorentz scalar Feynman integrals into a minimal Master Integral (MI) basis, dimensional shift reduction (DSRs)~\cite{Tarasov:1996br,Tarasov:1997kx} of divergent MIs in terms of finite MIs, and the direct integration of MIs into Generalized Polylogarithms (GPLs)~\cite{Goncharov:1998kja,Goncharov:2001iea,Panzer:2014caa}.
Together with other standard approaches, reviewed e.g. in Ref.~\cite{Heinrich:2020ybq}, these tools allowed the community to provide a great number of amplitude results relevant to Large Hadron Collider (LHC) phenomenology, see an example review in Ref.~\cite{Huss:2025nlt}.

In parallel to the approach presented in this paper, there have been other recent developments in understanding nonperturbative QCD effects.
One example is the estimation of subleading power corrections~\cite{Beneke:2021lkq,Caola:2021kzt} in the factorization theorem for the hadronic cross section using renormalons~\cite{Altarelli:1995kz,Beneke:1998ui}.
More formally, the S-matrix bootstrap for the nonperturbative four-pion scattering amplitude has been investigated~\cite{Guerrieri:2018uew}.
Another bootstrap method was applied to a nonperturbative four-scalar amplitude using consistency constraints such as positivity, analyticity, crossing, and unitarity~\cite{Tourkine:2023xtu,Albert:2023seb}.
In addition, a lot of progress has been made in understanding the structure of Feynman integrals beyond the fixed-order formulation~\cite{Radozycki:1997dn,Broadhurst:1999ys,Weinzierl:2015nda,Cavalcanti:2020osb,Brown:2024yvt,Borinsky:2025stz,Balduf:2025idm,Clingerman:2025pdy,Lippstreu:2025jit}.

The rest of this paper is organized as follows.
In sec.~\eqref{sec:back}, we review a theoretical background on the BSE and DSEs.
In sec.~\eqref{sec:alt}, we describe an alternative path towards solving the DSEs analytically starting from an all-loop perturbative analysis.
In secs~\eqref{sec:gg},~\eqref{sec:qq},~\eqref{sec:BSker}, we detail our calculation and its result for the gluon propagator, quark propagator, and BS kernel, respectively, in massless subleading large number of flavors $N_f$ approximation of QCD.
In sec.~\eqref{sec:concl}, we present our conclusions and discuss possible future directions.

\section{Background}
\label{sec:back}

In this section, we give a brief introduction to the Bethe-Salpeter and Dyson-Schwinger equations that motivate our calculation presented later in this paper.
We compare their properties in full QCD and in the 't~Hooft model~\cite{tHooft:1974pnl}.

\subsection{Bethe-Salpeter equation for the mesonic spectrum}

Consider a process involving a quark-antiquark pair of possibly different flavors $f\bar{f'}$ and a pseudoscalar meson $\phi$
\begin{equation}
f(-q_1) + \bar{f'}(-q_2) \to \phi(q_3) \,.
\label{eq:ffm}
\end{equation}
We choose an all-outgoing convention for the four-momenta $q_i^\mu$ so that the corresponding momentum conservation reads $q_1+q_2+q_3=0$.
In general, we do not assume on-shell conditions on the external quarks with masses $m_1$ and $m_2$, i.e. their momenta can be off-shell $q_1^2 \neq m_1^2 \,$ and $q_2^2 \neq m_2^2 \,$.
The meson-quark-antiquark correlator $\Gamma(q_1,q_2)$ satisfies the BSE
\begin{equation}
\Gamma(q_1,q_2) = \int \frac{d^d k}{(2\pi)^d} \, S(k+q_{12}) \, \Gamma(k+q_{12},-k) \, S(k) \, K(q_1,q_2,k)
\label{eq:BS}
\end{equation}
\centerline{\includegraphics[width=0.5\textwidth]{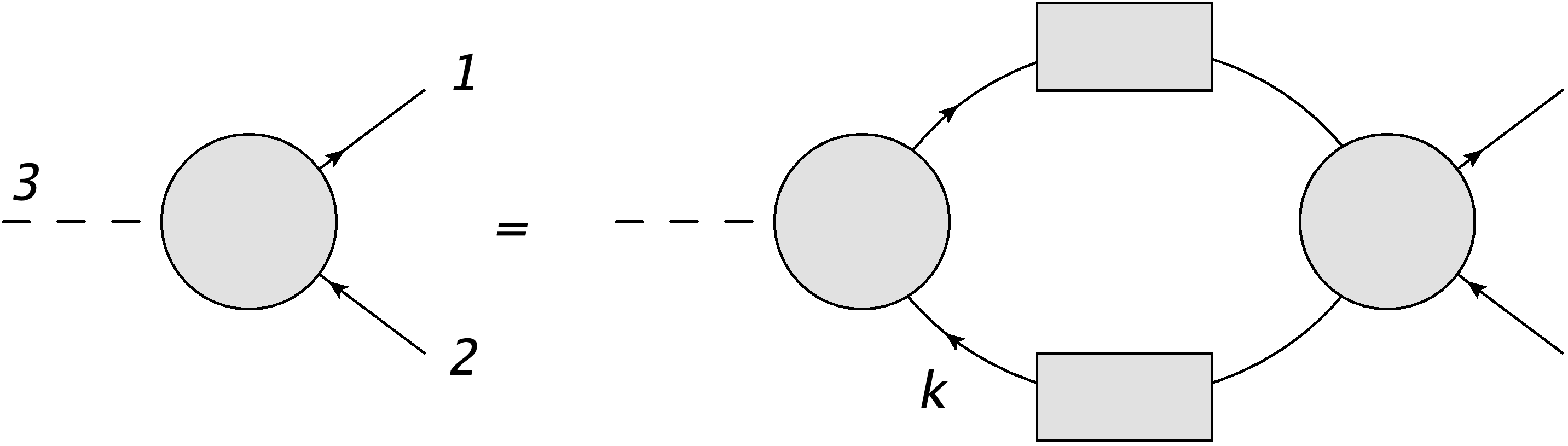}}
for a given full quark propagator $S(k)$ with loop momentum $k$ flowing along the spinor arrow, and the BS kernel $K(q_1,q_2,k)$ which describes the interactions among the four quarks.
Throughout this work, we denote a sum of all one-particle-irreducible (1PI) diagrams by a round blob, and a sum of all both reducible and irreducible diagrams by a square blob, and we refer to the corresponding correlators as 1PI and full, respectively. 
We also refer to the perturbative leading order undressed correlator as free and denote it by subscript 0.
Note that the order of the functions in the integrand is important because they contain noncommutative Dirac gamma matrices
\begin{equation}
\begin{split}
\Gamma(q_1,q_2) &= \gamma^5(\Gamma_5 \, \mathbb{1} + \Gamma_1 \, \slashed{q}_1 + \Gamma_2 \, \slashed{q}_2 + \Gamma_{12} \, \slashed{q}_1 \slashed{q}_2) \,, \\
S(k) &= S_m \, \mathbb{1} + S_k \, \slashed{k} \,, \\
K &= \sum_t K_t \, T_t \,,
\end{split}
\label{eq:introTensors}
\end{equation}
where the Lorentz scalar functions depend on the following scalars $\Gamma_i(q_1^2,q_2^2,q_3^2,m_1^2,m_2^2)$ and $S_i(k^2,m_k^2)$, while we elaborate on the matrices $T_t$ and form factors $K_t$ of the BS kernel $K$ later in sec.~\eqref{sec:BSker.gen}.
For given full nonperturbative fermion propagator $S$ and BS kernel $K$, the BSE~\eqref{eq:BS} is valid nonperturbatively both in the strong and weak coupling region, i.e. at high and low energies, respectively.
It describes how the mesonic bound state forms from the valence quark pair. 
For a given quark pair kinematics $q_1^2$, $q_2^2$, $m_1^2$ and $m_2^2$, the solutions to the BSE~\eqref{eq:BS} are eigenfunctions $\Gamma^{(n)}$ with eigenvalues $q_3^2=(m_3^{(n)})^2$.
As such, the BSE~\eqref{eq:BS} explicitly provides the spectrum of allowed masses of mesons $m_3^{(n)}$.

In the 't~Hooft model, the mesonic mass spectrum is discrete and its Regge trajectories are nearly linear in $n$.
It also exhibits a mass gap, i.e. $m_3^{(0)}>0$.
From experimental observations and numerical predictions, we expect the full QCD hadronic spectrum to also be discrete and exhibit a mass gap.
Note that in the 't~Hooft model, the Coulomb potential is linear $V_{\text{2D}} \sim r$, therefore, a free gluon propagator in the BS kernel is already strong enough to form be confining.
In comparison, in full QCD, the Coulomb potential is much weaker $V_{\text{4D}} \sim -\frac{1}{r^2}$, therefore both quark and gluon propagators need to be dressed with quantum corrections.
As a result, they numerically undergo a dynamical mass generation through dynamical chiral symmetry breaking.
This mechanism makes the analytic predictions much more involved in QCD.

In QCD, hadrons are bound states of $N_f=5$ valence light quarks.
The Particle Data Group (PDG)~\cite{ParticleDataGroup:2020ssz} experimental values of the masses of these quarks read
\begin{equation}
\begin{split}
m_u &= 2.16 \pm 0.04 \, \text{MeV} \,, \quad
m_s = 93.5 \pm 0.5 \, \text{MeV} \,, \quad \\
m_d &= 4.70 \pm 0.04 \, \text{MeV} \,, \quad
m_c = 1273.0 \pm 2.8 \, \text{MeV} \,, \quad
m_b = 4183 \pm 4 \, \text{MeV} \,,
\end{split}
\end{equation}
while the remaining sixth up-type heavy quark of mass $m_{t,\text{pole}} = 172.4 \pm 0.7 \, \text{GeV}$ decays before hadronization.
The masses of the hadrons depend on their valence quark content and the remaining binding energy
\begin{equation}
m_\phi = m_f + m_{\bar{f'}} + \Delta E_\phi \,.
\label{eq:mphi}
\end{equation}
For example, the pseudoscalar mesonic spectrum in the unflavored sector starts with pion
\begin{equation}
\pi^0 = \left< \frac{u\bar{u}-d\bar{d}}{\sqrt{2}} \right> \,,
\qquad 
m_{\pi^0} = 134.9768 \pm0.0005 \, \text{MeV} \,,
\label{eq:mpi}
\end{equation}
while in the bottomonium sector, with the bottom eta meson
\begin{equation}
\eta_b = \left< b\bar{b} \right> \,,
\qquad 
m_{\eta_b} = 9.3987 \pm 0.002 \, \text{GeV} \,.
\label{eq:meta}
\end{equation}
Note that the binding energy constitutes a larger portion of the mesonic mass for lighter bound states, e.g. $\frac{\Delta E_{\pi^+}}{m_{\pi^+}} = 95\%$, than for heavier ones, e.g. $\frac{\Delta E_{\eta_b}}{m_{\eta_b}} = 11\% \,$.

Before concluding the discussion on the BSE~\eqref{eq:BS}, it is worth pointing out that it is possible to transform the BSE into a form which does not require integrating over the two full quark propagators $S(k)$ and $S(k+q_{12})$.
Indeed, one could dress the 1PI correlator $\Gamma$ with the two quark propagators
\begin{equation}
\tilde\Gamma(q_1,q_2) = S(q_1) \, \Gamma(q_1,q_2) \, S(-q_2) \,,
\end{equation}
such that $\tilde\Gamma$ is no longer 1PI.
After substituting into BSE~\eqref{eq:BS} and multiplying by $S(q_1)$ from the left-hand side (LHS) and by $S(-q_2)$ from the right-hand side (RHS), we obtain
\begin{equation}
\tilde\Gamma(q_1,q_2) = \int \frac{d^d k}{(2\pi)^d} \, S(q_1) \, \tilde\Gamma(k+q_{12},-k) \, S(-q_2) \, K(q_1,q_2,k) \,.
\end{equation}
\centerline{\includegraphics[width=0.6\textwidth]{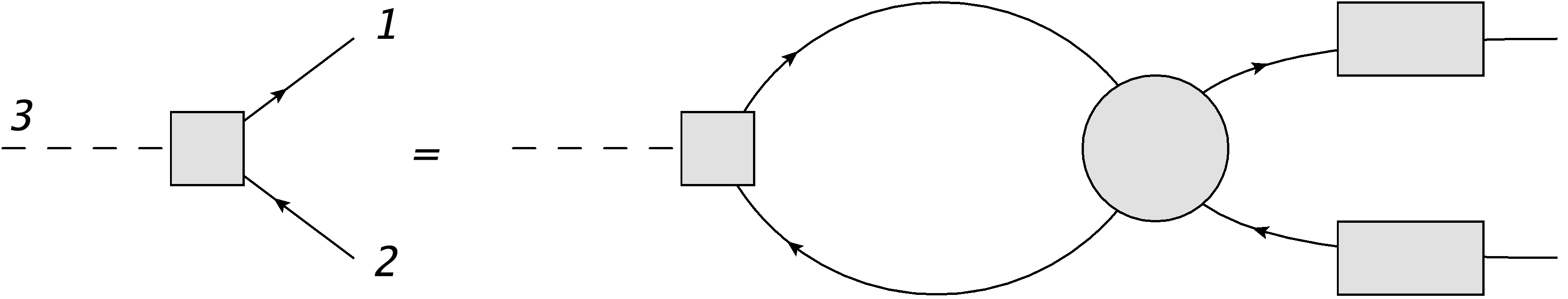}}
Note that the quark propagators no longer depend on the loop momentum $k$.

\subsection{Dyson-Schwinger equations}

The BSE~\eqref{eq:BS} for the mesonic spectrum requires the full quark propagator $S$ and the BS kernel $K$ as known inputs.
These correlators satisfy the DSEs
\begin{equation}
\frac{1}{i} S(p)^{-1} = \frac{1}{i} S_0(p)^{-1} + \int \frac{d^d k}{(2\pi)^d} \, \frac{1}{i} D^{\mu \nu a b}(k-p) \, (- i g \, \gamma_\nu \, T^b) \, i \, S(k) \, V_\mu^a(p,k) \,,
\label{eq:DS}
\end{equation}
\centerline{\includegraphics[width=0.7\textwidth]{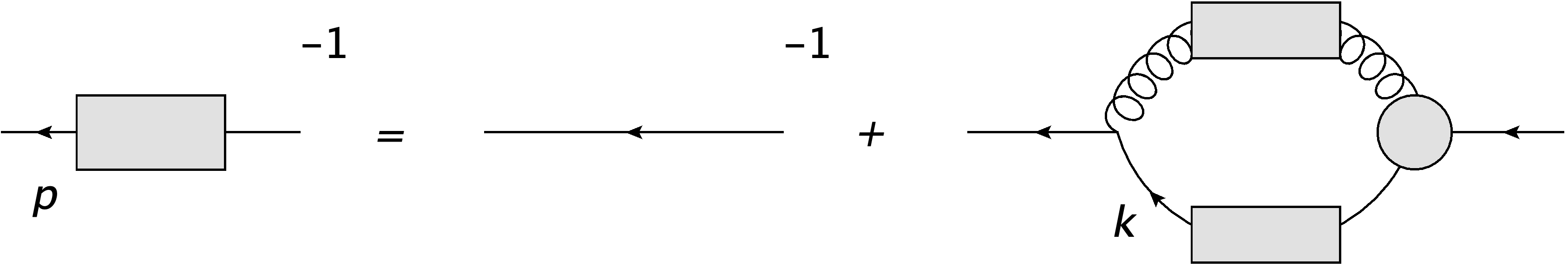}}
\begin{equation}
\vdots \notag
\end{equation}
where $S_0^{-1} = \slashed{p} - m$ is the inverse free quark propagator, $D^{\mu \nu a b}$ is the full gluon propagator with Lorentz indices $\mu,\nu$ and adjoint color indices $a,b$, $g$ is the QCD coupling, $T^b$ is the fundamental generator of SU($N_c$), and $V_\mu^a$ is the 1PI gluon-quark-antiquark vertex.
Similarly to the BSE~\eqref{eq:BS}, the DSEs are valid nonperturbatively and can be derived using functional methods from the path integral formulation of QCD~\cite{Swanson:2010pw}.
Importantly, the DSEs form an infinite coupled system, which we denote by the dots.
Indeed, the quark two-point function $S$ depends on the three-point function $V$, which depends on four-point functions of QCD particles in further equations, and so on.
This makes it prohibitive to solve the DSEs directly, even for the lower-point functions, as they are fully coupled integral equations.

Similarly to the BSE~\eqref{eq:BS}, the DSEs~\eqref{eq:DS} are usually possible to solve only numerically and under certain assumptions.
In practice, an infinite system of the DSEs has to be truncated to make solving it feasible.
In the resulting finite system of coupled equations for the lower-point functions, the higher-point functions are modeled separately.
However, since there is no parameter that controls the precision of a specific truncation choice, truncating the DSEs is not a rigorous approximation but rather an ambiguous scheme.
This poses a great theoretical challenge, which we address in this work.
Recently, this problem has been tackled e.g. in Refs~\cite{Binosi:2007pi,Binosi:2016rxz,Banks:2024ydh,Peng:2024azv}.

One of the simplest truncation schemes is the so called rainbow-ladder scheme, which is exact in the 't~Hooft model.
In this scheme, one treats the full gluon propagator $D^{\mu \nu a b}$, the 1PI vertex $V_\mu^a$ and the BS kernel $K$ as their free counterparts $D_0^{\mu \nu a b}$, $V_{0,\mu}^a$ and $K_0$, respectively.
As a result, the BSE~\eqref{eq:BS} and the DSEs~\eqref{eq:DS} simplify significantly.
Indeed, the infinite system of DSEs~\eqref{eq:DS} decouples, so that we are left with a single DSE for the quark propagator, and the usual one BSE for the mesonic mass gap
\\
\centerline{\includegraphics[width=0.7\textwidth]{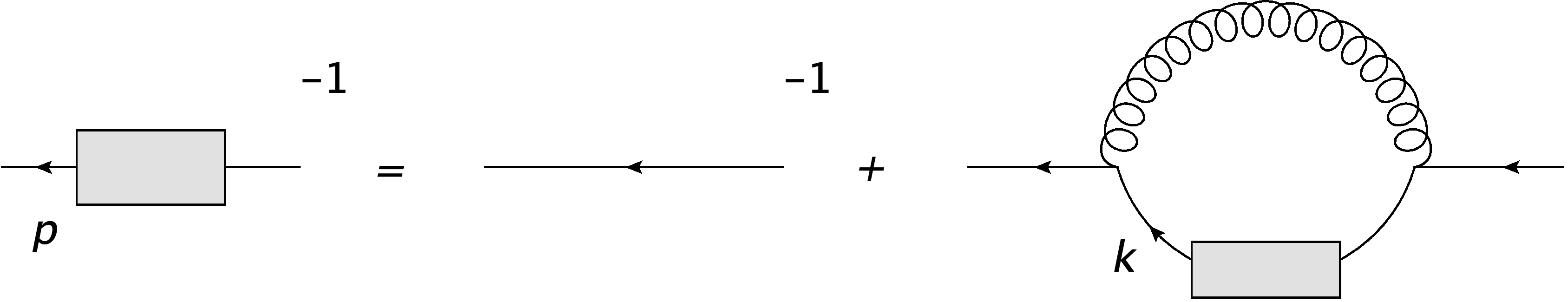}}
\centerline{\includegraphics[width=0.5\textwidth]{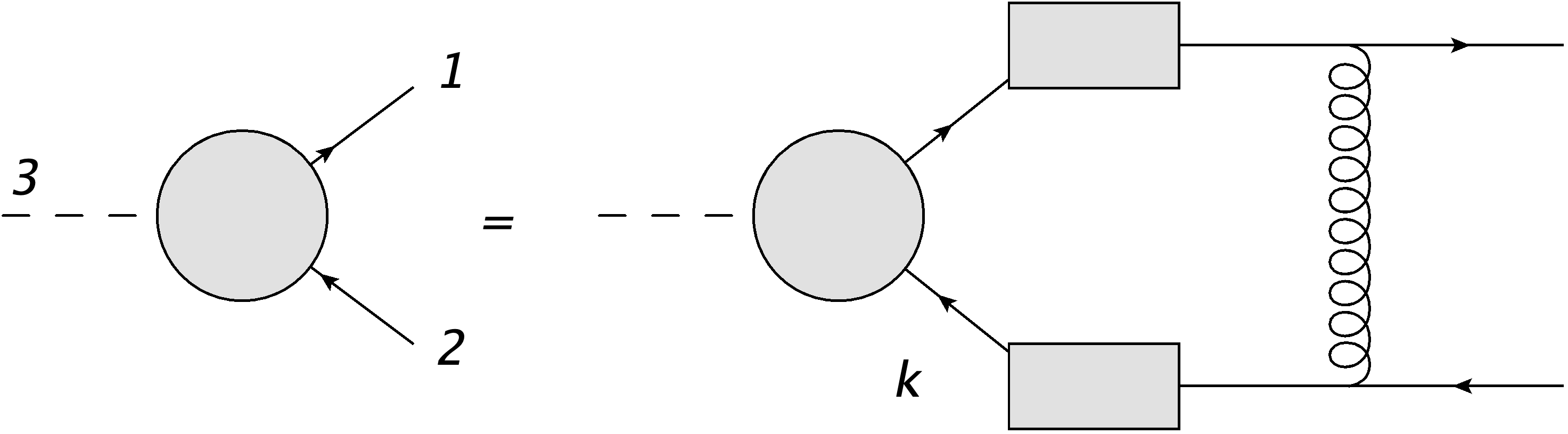}}
\noindent
Amazingly, 't~Hooft was able to exactly solve this DSE in $d_0=2$ for the full quark propagator $S$, and then analytically estimate the mass gap arising from the corresponding BSE.
However, in full QCD, the rainbow-ladder scheme is no longer exact.
In modern applications, a scheme that originates in the skeleton expansion is used.
In this scheme, one keeps all the DSEs~\eqref{eq:DS} corresponding to the correlators appearing explicitly in QCD Lagrangian, i.e. quark, gluon and ghost propagators, as well as three-gluon, gluon-quark-antiquark, gluon-ghost-ghost and four-gluon vertices.
Remarkably, such a scheme provides results that are close to those of LQCD, suggesting that the contribution of the higher-point vertices can be somehow suppressed~\cite{Alkofer:2000wg,Fischer:2006ub}.

When the value of the QCD coupling $g$ is small, the recursive DSEs~\eqref{eq:DS} are equivalent to a perturbative expansion in higher-loop loop Feynman diagrams.
At a fixed perturbative order, the equations for higher-point functions that are inaccessible at this order decouple, so the truncation is controlled by the power of the QCD coupling $g$.
For example, consider a Feynman diagram contributing to the quark propagator
\\
\centerline{\includegraphics[width=0.4\textwidth]{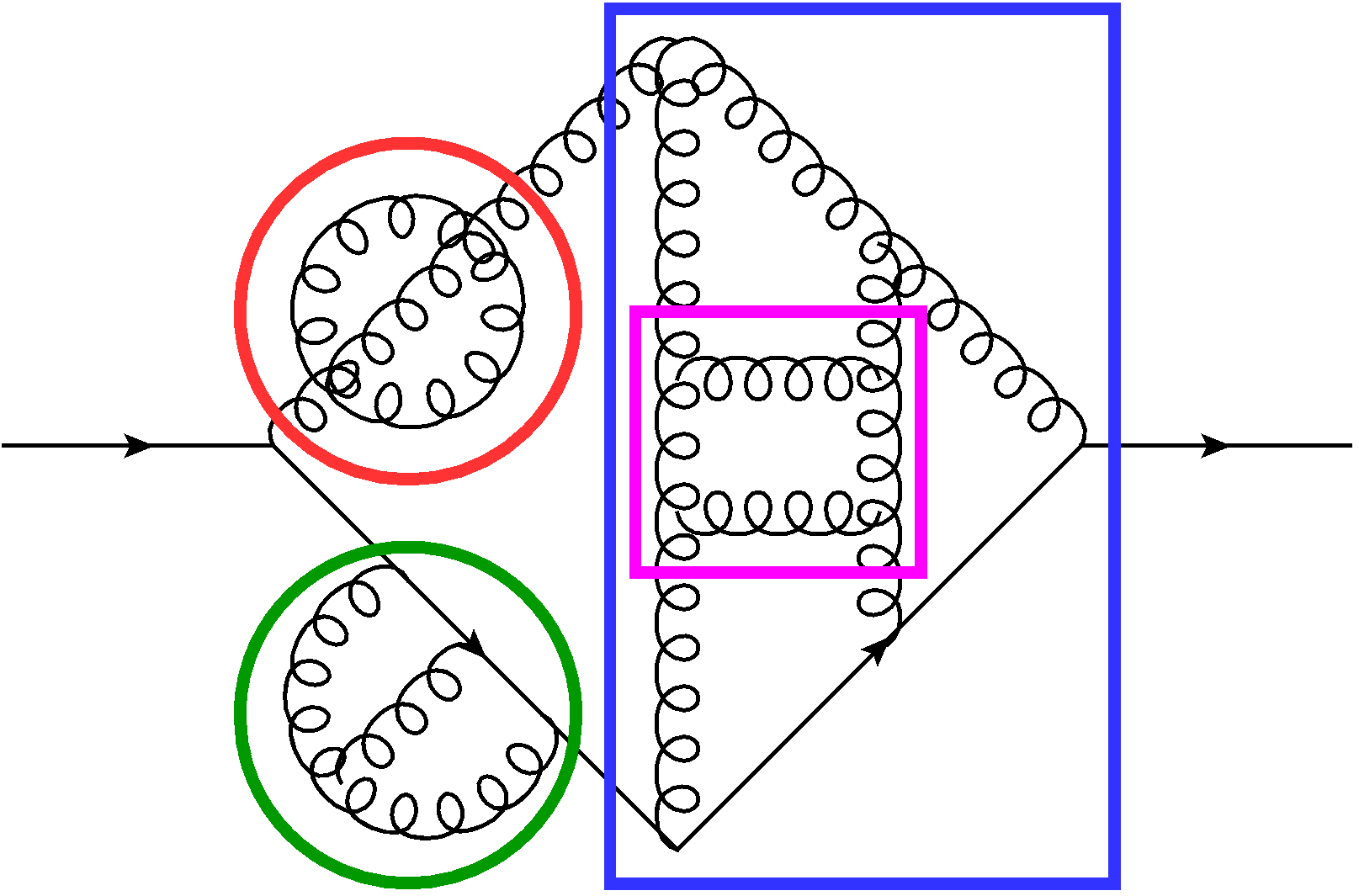}}
\noindent
It involves higher-loop corrections to the gluon and quark propagator, as well as to the gluon-quark-antiquark vertex, marked in red, green, and blue lines, respectively.
Note that there is a correction to the four-gluon vertex within the gluon-quark-antiquark vertex, marked in violet.
This approximation of the DSEs~\eqref{eq:DS} is the starting point for our further considerations.

\section{Alternative solution to the Dyson-Schwinger equations}
\label{sec:alt}

In this section, we outline an alternative way to solve the DSEs~\eqref{eq:DS}.
First, we expand the correlator in a transseries representation.
Then, we argue that its perturbative part is feasible to be computed to all loops in the massless subleading large $N_f$ limit as a systematic approximation of QCD.
We also comment on the comparison with the other well-known approximations.
Finally, we present a brief overview of the modern scattering amplitude methods that we exploit in our explicit calculation.

\subsection{Transseries representation}

In this paper, we make the first step towards extending the work of 't~Hooft on computing the QCD mesonic spectrum in $d_0=2$ dimensions~\cite{tHooft:1973alw} to $d_0=4$ dimensions.
This means finding the eigenvalues of the BSE~\eqref{eq:BS}.
To this end, we first need to calculate the full quark propagator $S$ and the BS kernel $K$.
At the same time, we want to avoid the ambiguities of choosing a truncation scheme in the DSEs~\eqref{eq:DS}, which these two correlators satisfy.
For this reason, we take an alternative approach to solving the the DSEs~\eqref{eq:DS} directly, which is motivated by the resurgence theory, see recent progress e.g. in Refs~\cite{Dorigoni:2014hea,Bellon:2014zxa,Laenen:2023hzu,Caron-Huot:2023tpw,Maiezza:2024nbx,Dunne:2025mye}.
Indeed, we split any correlator $\C$ into its perturbative part $\C_{\text{pert}}$, which can be calculated around $g=0$, and the remaining nonperturbative part $\C_{\text{nonpert}}$.
The transseries expansion of such correlator reads
\begin{equation}
\C
= \C_{\text{pert}} + \C_{\text{nonpert}}
= \sum_{n \geq 0} c_n^{(0)} g^n
+ \sum_i e^{-\frac{S_i}{g}} \sum_{n \geq 0} c_n^{(i)} g^n \,,
\end{equation}
where $S_i$ are the nontrivial nonperturbative saddles labeled by an index $i$, e.g. instantons, and $c_n^{(i)}$ are the perturbative coefficients around each saddle.
By a careful analysis of the singularities in the whole complex Borel plane of the perturbative part of the correlator $\C_{\text{pert}}$, one can retrieve $\C_{\text{nonpert}}$.
Therefore, we need to calculate the parturbative part of the full quark propagator $S$ and BS kernel $K$.
In this work, we explicitly provide these all-loop bare correlators in a systematic approximation detailed below, and we leave further steps for future investigations.

\subsection{Massless subleading large $N_f$ approximation}

Calculating the perturbative part of the quark propagator $S$ and the BS kernel $K$ to all loops in full QCD still remains prohibitive.
Therefore, we seek an approximation of full QCD that would simultaneously
\begin{enumerate}
\item make an all-loop calculation feasible;
\item be systematically improvable;
\item and preserve asymptotic freedom.
\end{enumerate}
We argue that an example of such is the massless subleading large $N_f$ approximation, with a diagonal CKM matrix.

We define the large number of massless quarks $N_f$ approximation via
\begin{equation}
N_f \to \infty \,, 
\quad \text{while} \quad 
g^2 N_f = \lambda \quad \text{fixed} \,,
\end{equation}
where we refer to $\lambda$ as the effective coupling.
Note that the second condition is necessary for the power series to be meaningful.
Indeed, in general, each new loop order in QCD introduces terms proportional to scalar color factors $\{N_f,C_A,C_F\}$,
where the quadratic Casimir invariants are
\begin{equation}
C_A = N_c = 3 \,,
\qquad
C_F = \frac{N_c^2-1}{2N_c} = 1.(3) \,.
\end{equation}
Therefore, we cannot simply expand in powers of $N_f$, but rather of $\lambda$.
To subleading order in large $N_f$, the corresponding perturbative series in $\lambda$ for any correlator $\C$ organizes as follows
\begin{equation}
\begin{split}
\C_{\text{pert}} 
=\,& g^2 ( \C_0 + g^2 \, \C_1 + g^4 \, \C_2 + \dots ) \\
=\,& g^2 ( \C_0 + g^2 (N_f \, \C_{1N} + C_A \, \C_{1A} + C_F \, \C_{1F} ) \\
&+ g^4 (N_f ( N_f \, \C_{2N} + C_A \, \C_{2A} + C_F \, \C_{2F}) + \mathcal{O}(N_f^0) ) + \dots ) \\
=\,& g^2 \left( \C_0 + \lambda \left( \C_{1N} + \frac{C_A}{N_f} \, \C_{1A} + \frac{C_F}{N_f} \, \C_{1F} \right) \right) \\
&+ g^2 \left( \lambda^2 \left( \C_{2N} + \frac{C_A}{N_f} \, \C_{2A} + \frac{C_F}{N_f} \, \C_{2F} \right) + \cdots + \mathcal{O}(N_f^{-2}) \right)
\end{split}
\end{equation}
where $\C_{Lc}$ denotes the corresponding correlator contribution at $L$ loops to the color factor $c$, the dots denote an infinite sum to all orders in the effective coupling $\lambda$, while $\mathcal{O}(N_f^{-2})$ denotes the sub-subleading corrections that we are going to suppress.

Let us comment on the properties of this approximation, which we required above.
\begin{enumerate}

\item 
At leading order in large $N_f$, the corresponding correction to a correlator consists of only one-loop-factorizable diagrams, i.e. higher-loop corrections arise from multiplying one-loop and tree-level diagrams.
As one-loop corrections are well-understood (see example review in Ref.~\cite{Heinrich:2020ybq}), this allows for all-order computations, as possible in e.g. QCD resummation (see e.g. Ref.~\cite{Catani:1989ne}).
In order to account for all loops in $\lambda$, one must consider up to an infinite chain of closed quark loops, i.e.
\\
\centerline{\includegraphics[width=0.5\textwidth]{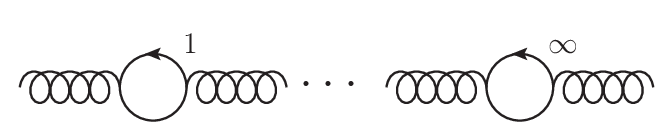}}
\noindent
At subleading order in large $N_f$, this infinite chain may appear inside another loop correction, e.g.
\\
\centerline{\includegraphics[width=0.5\textwidth]{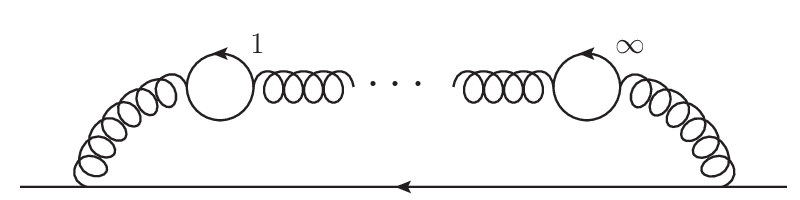}}
\noindent
We refer to the loop order that underlies the infinite bubble chain as the base loop order, e.g. the above diagram is one-loop-based.
As the analytic expression for the one-loop bubble in the quark chain is well-known, the resulting increase in the required computational complexity in comparison to the base order is manageable.
Importantly, for each process, there is only a finite number of base loop orders in this approximation, i.e. the higher-point DSEs decouple, as expected from the truncated results discussed in sec.~\eqref{sec:back}.
For example, for the BS kernel correlator, we have to consider up to three-loop-based diagrams at subleading large $N_f$.
Importantly, the applicability of modern scattering amplitude methods makes the all-loop calculation feasible.

\item 
As the large $N_f$ limit provides a self-consistent power expansion, the precision of the prediction made in this approximation can be improved systematically by considering further sub$^n$-leading corrections in $\frac{1}{N_f}$.
Note that the valence quarks along the external lines of the $BS$ kernel and the sea quarks contributing to the loop corrections could be treated separately in different approximations.
Indeed, one could improve the precision by accounting for the masses of either valence or sea quarks, or both. 
However, we leave these improvements for future work.

\item 
It is well-established that massless QCD is asymptotically free when its one-loop coefficient of the beta function is negative~\cite{Gross:1973id,Politzer:1973fx}, i.e.
\begin{equation}
\beta_0 = -\left( \frac{11}{3}C_A - \frac{2}{3}N_f \right) < 0 \quad \text{for} \quad N_f < 5.5 \, C_A\,.
\label{eq:beta}
\end{equation}
Note that this coefficient in the large $N_f$ approximation to subleading order stays exactly the same.
Therefore, our approximation preserves asymptotic freedom as long as $N_f<5.5 \, C_A$.
Note that the further $N_f$ is from 0, and the closer it is to $5.5 \, C_A$, the worse is the suppression of nonperturbative effects~\cite{Gardi:1998ch}.
On the other hand, we also need to guarantee that the other color factors generated by each loop are smaller than $N_f$, i.e.
\begin{equation}
C_F,C_A < N_f \,.
\end{equation}
Since we have $C_F<C_A$ in SU(3), we finally obtain the following validity bounds on the value of $N_f$
\begin{equation}
3 = C_A < N_f < 5.5 \, C_A = 16.5 \,.
\end{equation}

\end{enumerate}

Let us also elaborate on the phenomenological relevance of the massless quark approximation.
It is valid when the experimental values of the masses of all considered valence $m_f,m_{\bar{f'}}$ (as in eq.~\eqref{eq:mphi}) and sea $m_i$ quarks are much smaller than the characteristic scale in the process, i.e.
\begin{equation}
m_f,m_{\bar{f'}},m_i \ll m_\phi \,, \qquad i=1,\dots,N_f \,.
\end{equation}
Note that $m_f,m_{\bar{f'}} \ll m_\phi$ is valid only for mesons that involve valence quarks belonging to any subset of the three lightest quarks, i.e. up, down, and strange, see sec.~\eqref{sec:back}.
In the approximation considered here, both the valence and sea quarks are treated as massless, thus we expect the meson mass to be of order of the pion mass~\eqref{eq:mpi}.
We point out that for $N_f>3$ sea quarks required above, the resulting meson mass will receive additional contribution arising from the $(N_f-3)$ new unphysical light sea quarks, as opposed to the physical suppression of the contributions stemming from the heavy sea quarks, i.e. charm and bottom.
Note that if all the considered quarks are massless, then the meson mass for any different valence constituents will be the same.
In order to distinguish between mesons made of different flavors of their valence quarks, one would have to account for their masses.
This approximation would be valid e.g. for bottomonium with $N_f=4$ massless sea quarks.
Similarly, keeping also the experimental values of the masses of the sea quarks, would be valid for bottomonium with $N_f=4$ or $N_f=5$.
It is worth pointing out that the more mass scales are included in the calculation, the more complex it becomes.

\subsection{Comparison with other approximations}

Let us briefly comment here on the comparison of our subleading large $N_f$ approximation with other well-established systematic approximations.

First, consider the leading large $N_f$ approximation.
This approximation is important e.g. for the renormalon contributions to QCD~\cite{Altarelli:1995kz,Beneke:1998ui}
Although it makes an all-loop calculation easier than in our subleading approximation, the resulting theory is no longer asymptotically free.
Indeed, the one-loop coefficient of the corresponding beta function is positive
\begin{equation}
\beta_0 = \frac{2}{3}N_f >0 \,.
\end{equation}
For this reason, we need to consider at least a subleading order correction in this limit.

Second, consider the large $N_c$ approximation.
It is defined by~\cite{tHooft:1973alw}
\begin{equation}
N_c \to \infty \,, 
\quad \text{while} \quad 
g^2 N_c = \text{fixed} \,.
\end{equation}
The resulting theory is asymptotically free, i.e.
\begin{equation}
\beta_0 = -\frac{11}{3}C_A <0 \,.
\end{equation}
At leading order in large $N_c$, the planar gluonic corrections dominate in any process that involves only colored states.
In addition, the pure Yang-Mills theory becomes semi-classical, i.e. the perturbative part of the correlator dominates over the nonperturbative instantonic part suppressed~\footnote{We note that other nonperturbative effects might not be fully suppressed, see e.g. Ref.~\cite{Marino:2012zq}.} as $e^{-N_c}$.
This approximation has been successfully used throughout the years in e.g. finding the mesonic spectrum of the 't~Hooft model~\cite{tHooft:1974pnl}, establishing the baryon mass scaling by Witten~\cite{Witten:1979kh}, introducing the AdS/CFT correspondence between String and super-Yang-Mills theory~\cite{Maldacena:1997re,Witten:1998qj}.
However, each additional loop correction introduces contributions from higher-point correlators.
For example, a twenty-four-loop gluonic fishnet Feynman diagram for the quark propagator
\\
\centerline{\includegraphics[width=0.4\textwidth]{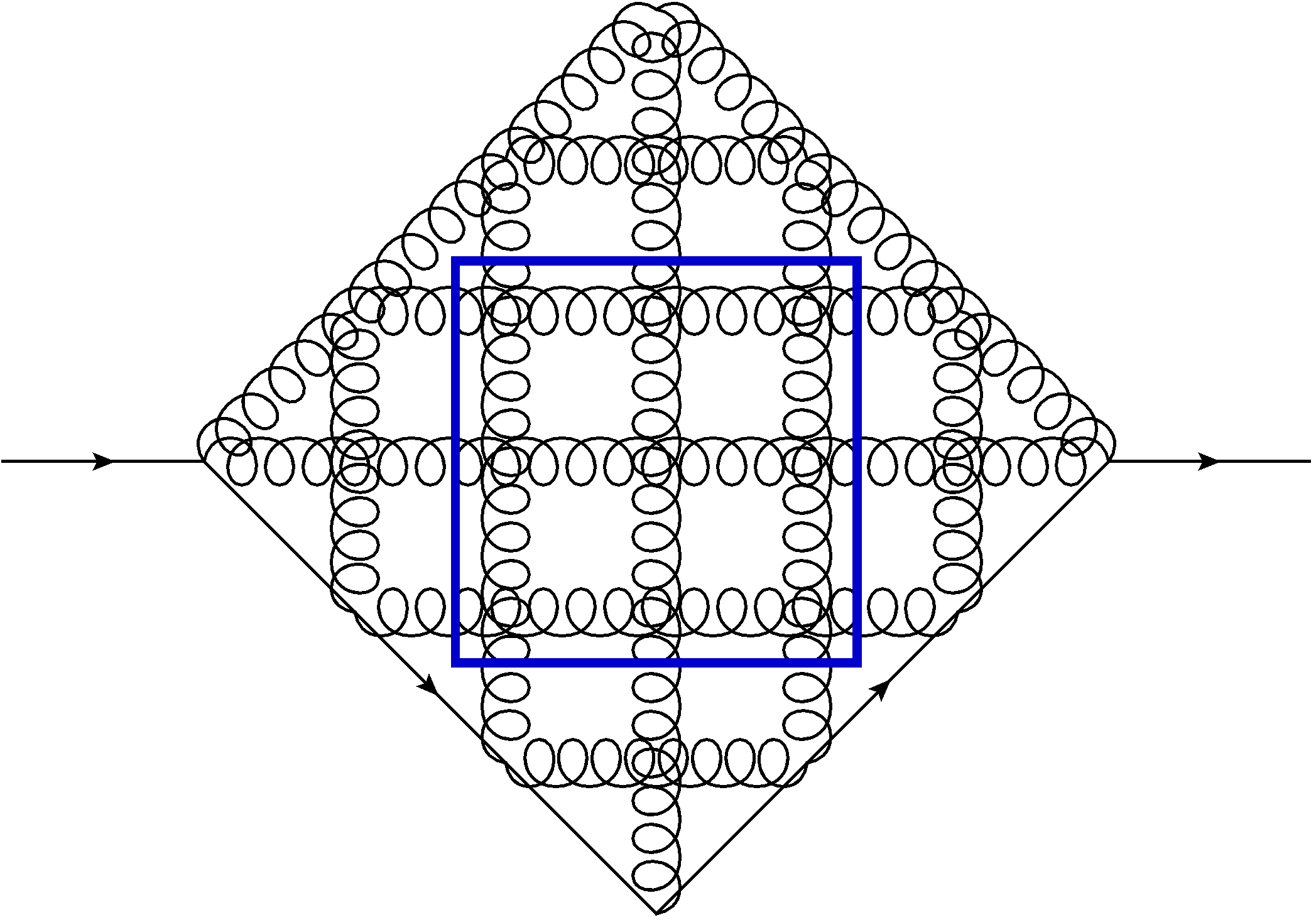}}
\noindent
receives a contribution from a four-loop twelve-gluon correlator.
Since the all-loop calculation in this approximation is not feasible at the moment, we do not consider it further in this work.

Finally, let us mention that the Veneziano approximation~\cite{Veneziano:1976wm} allows for a simultaneous large $N_c$ and $N_f$ expansion.
It is defined via
\begin{equation}
N_c \,, N_f \to \infty \,, \quad \text{while} \quad g^2 N_c \,, \frac{N_f}{N_c} \quad  \text{fixed} \,.
\end{equation}
Similarly to the large $N_c$ limit, the planar gluonic diagrams are dominant, while the quark corrections are suppressed.
Unfortunately, the all-loop calculation is also not manageable.

\subsection{Scattering amplitude methods}

Before presenting the details of our calculation of the aforementioned correlators, let us elaborate on the methods that we used throughout this work.
They follow a standard powerful computational amplitude workflow, that was applied successfully over the last years~\cite{Heinrich:2020ybq,Huss:2025nlt}.

We perform all the calculations in dimReg around
\begin{equation}
	d=d_0-2\ep \qquad \text{with} \qquad d_0=4 \qquad \text{and} \qquad \ep \to 0 \,.
\end{equation}
For convenience, we choose the Landau gauge fixing
\begin{equation}
	\xi = 0 \,.
\end{equation}
Note that since the large $N_f$ approximation is gauge-invariant order by order, the resulting physical quantities do not depend on the gauge choice.
We define a correlator $\C$ for a scattering process as a sum of corresponding Feynman diagrams
\begin{equation}
	\C_{\text{pert}} = \sum \text{Feynman diagrams} \,.
\end{equation}
We generated all the Feynman diagrams with \texttt{qgraf}~\cite{Nogueira:1991ex}.
By performing the SU(3) color, Lorentz tensor, and Dirac spinor algebra in \texttt{FORM}~\cite{Vermaseren:2000nd}, we decompose the correlator $\C_{\text{pert}}$ into a linear combination of color tensors $\mathcal{C}_c$, Lorentz and Dirac tensors $T_t$, and scalar Feynamn integrals $I_{\vec{n}}$
\begin{equation}
	\C_{\text{pert}} = \sum_{c,t,\vec{n}} r_{c,t,\vec{n}} \, \mathcal{C}_c \, T_t \, I_{\vec{n}} \,,
\end{equation}
where $r_{c,t,\vec{n}}$ are rational functions of the dimension and kinematic invariants.
All the resulting $L$-loop Feynman integrals $I_{\vec{n}}$ can be classified by the underlying integral topology (topo), which is defined by a set of $N$ generalized propagators $\{\mathcal{D}_i^{(\text{topo})}\}$, and a set of the corresponding integer propagator powers $\vec{n}$
\begin{equation}
I^{(\text{topo})}_{\vec{n}} 
= \int \left( \prod_{l=1}^{L} \frac{d^d k_l}{(2\pi)^d} \right)
\frac{1}{(\mathcal{D}_1^{(\text{topo})})^{n_1} \cdots (\mathcal{D}_N^{(\text{topo})})^{n_N}} \,,
\end{equation}
where $k_l$ are the $d$-dimesional loop momenta.
The scalar Feynman integrals satisfy the so called Integration-By-Parts relations~\cite{Tkachov:1981wb,Chetyrkin:1981qh}
\begin{equation}
\int \left( \prod_{l=1}^{L} \frac{d^d k_l}{(2\pi)^d} \right)
\frac{\partial}{\partial k_i^\mu} 
\frac{q^\mu}{(\mathcal{D}_1^{(\text{topo})})^{n_1} \cdots (\mathcal{D}_N^{(\text{topo})})^{n_N}}
= 0
\end{equation}
for any vector $q^\mu$ and loop momentum $k_i^\mu$.
These relations allow us to perform a linear reduction to a minimal set of basis integrals called Master Integrals $M_i^{(\text{topo})}$
\begin{equation}
I^{(\text{topo})}_{\vec{n}}
= \sum_i c^{(\text{topo})}_{\vec{n},i} \, M_i^{(\text{topo})} \,,
\end{equation}
where $c^{(\text{topo})}_{\vec{n},i}$ are rational functions of the dimension and kinematic invariants.
Due to the presence of symbolic powers in the correlators considered below, we performed the required IBP reductions using \texttt{Kira3}~\cite{Lange:2025fba}.
Note that keeping a different subset of powers symbolic may result in a different number of MIs within the same topology definition.
In addition, it is sometimes beneficial to choose particular MIs such that they are finite in the physical limit $\ep \to 0$.
We can achieve this by increasing the dimension $d$ in the loop momentum integration.
To this end, we exploited the so called dimensional shift relations~\footnote{Note that the we perform the dimension shift in $d^d k/(i\pi^{d/2})$ integration measure instead of the standard $d^d k/(2\pi)^d$ one, to avoid spurious powers of $4\pi$.}~\cite{Tarasov:1996br,Tarasov:1997kx} with \texttt{LiteRed}~\cite{Lee:2013mka}.
In combination with the IBP reduction, any dimension-shifted integral can be decomposed to the original unshifted set of MIs
\begin{equation}
I^{(\text{topo},d+2)}_{\vec{n}}
= \sum_i c_{\vec{n},i}'^{(\text{topo})} \, M_i^{(\text{topo})} \,,
\end{equation}
thus allowing for a change of the MI basis.
All the MIs considered here can be expanded in powers of the regulator $\ep$
\begin{equation}
M_i^{(\text{topo})} 
= \sum_{j,\alpha} \ep^{j} \, a_{i,j,\alpha}^{(\text{topo})} \, G_\alpha^{(\text{topo})} 
\end{equation}
with coefficients $a_{i,j,\alpha}^{(\text{topo})}$ that depend on kinematics.
The transcendental functions of kinematics $G_\alpha^{(\text{topo})}$ are Generalized Polylogarithms~\cite{Goncharov:1998kja,Goncharov:2001iea} defined by
\begin{equation}
G(\alpha_n,\dots,\alpha_1;x) = \int_0^x \frac{dz}{z-\alpha_n} G(\alpha_{n-1},\dots,\alpha_1;z)\,,\,\,
G(\underbrace{0,\dots,0}_{n};x) \equiv \frac{\ln^n x }{n!} \,,\,\,
G(x)=1 \,.
\end{equation}
We evaluated the MIs in terms of GPLs by hand for bubble-factorizable ones, and using \texttt{HyperInt}~\cite{Panzer:2014caa} for the remaining ones.
We used \texttt{HypExp}~\cite{Huber:2005yg} to expand closed form MIs in terms of GPLs, \texttt{PolyLogTools}~\cite{Duhr:2019tlz} to manipulate GPLs, and \texttt{MultivariateApart}~\cite{Heller:2021qkz} to partial fraction the rational function coefficients.
In order to numerically double check our integral results, we used \texttt{AMFlow}~\cite{Liu:2022chg}.

\section{Gluon propagator}
\label{sec:gg}

We now proceed to the calculation of the bare full quark propagator $S$ and the BS kernel $K$ to all loops in the massless subleading large $N_f$ approximation of QCD.
As the full gluon propagator $D$ contributes to both, we start our analysis with it.

\subsection{General form}
\label{sec:gg.gen}

Let us start by describing the general form of a gluon propagator.
In the Landau gauge, the free gluon propagator $D_0$ at momentum $p$ reads
\begin{equation}
    \includegraphics[width=0.18\textwidth]{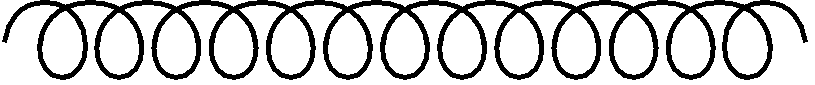} 
    = \frac{1}{i} D_0^{\mu\nu ab}(p) 
    = \delta^{ab} P^{\mu\nu} \frac{1}{i \, p^2}
\end{equation}
where $\mu,\nu=1,...,d$ are the Lorentz indices, and $a,b=1,...,N_c^2-1$ are the adjoint color indices.
We define a Lorentz tensor projector
\begin{equation}
    P^{\mu\nu}
    = g^{\mu\nu} - \frac{p^\mu p^\nu}{p^2}
\end{equation}
which satisfies the defining property
\begin{equation}
    P^\mu_\rho P^{\rho\nu} = P^{\mu\nu} \,,
\end{equation}
and its trace yields
\begin{equation}
    P^\mu_\mu = d-1 \,.
\end{equation}
The beyond-tree-level 1PI gluon propagator $\Pi$ has the following tensor form
\begin{equation}
    \includegraphics[width=0.1\textwidth]{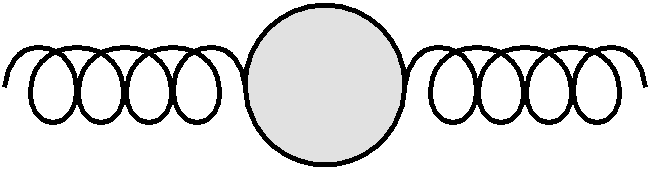}
    = i \, \Pi^{\mu\nu ab}(p) 
    = \delta^{ab} P^{\mu\nu} \, i \, \Pi(p^2) \,.
\label{eq:ggTen}
\end{equation}
which is universal for all the corrections considered below.
The scalar form factor can be defined by a projection
\begin{equation}
\Pi(p^2) 
= \frac{\delta^{ab}}{N_c^2-1} \frac{P_{\mu\nu}}{d-1} \Pi^{\mu\nu ab}(p) \,.
\label{eq:proj}
\end{equation}
The full gluon propagator $D$ results from summing up all the 1PI contributions $\Pi$ in a geometric series 
\\
\includegraphics[width=0.8\textwidth]{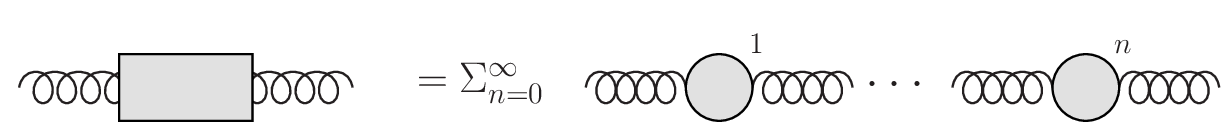}
\begin{equation}
\begin{split}
    \frac{1}{i} D^{\mu\nu ab}(p)
    &= \frac{1}{i} D_0^{\mu\nu ab}(p) + \frac{1}{i} D_0^{\mu\rho ac}(p) \, i \, \Pi_{\rho\sigma cd}(p) \, \frac{1}{i} D_0^{\sigma\nu db}(p) + ... \\
    &= \frac{\delta^{ab}}{i} \frac{P^{\mu\nu}}{p^2-\Pi(p^2)} \,.
\end{split}
\end{equation}

In the large $N_f$ approximation, we split the 1PI gluon propagator $\Pi$ into leading $(L)$ and subleading $(S)$ terms
\begin{equation}
    \Pi = \Pi_L + N_f^{-1} \, \Pi_S  + \mathcal{O}(N_f^{-2}) \,.
\end{equation}
The corresponding full gluon propagator $D$ reads
\begin{equation}
\begin{split}
    D^{\mu\nu ab}(p) 
    &= D^{\mu\nu ab}_L(p) + N_f^{-1} D^{\mu\nu ab}_S(p) + \mathcal{O}(N_f^{-2}) \\
    &= \delta^{ab} \, \frac{P^{\mu\nu}}{p^2-\Pi_L(p^2) -N_f^{-1} \Pi_S(p^2)} + \mathcal{O}(N_f^{-2})
\end{split}
\end{equation}
to subleading order in large $N_f$.
Let us now discuss the leading and subleading corrections, respectively.

\subsection{Leading order}

At leading order in large $N_f$, the all-loop 1PI gluon propagator is simply a one-loop correction stemming from each massless quark flavor
\\
\centerline{\includegraphics[width=0.5\textwidth]{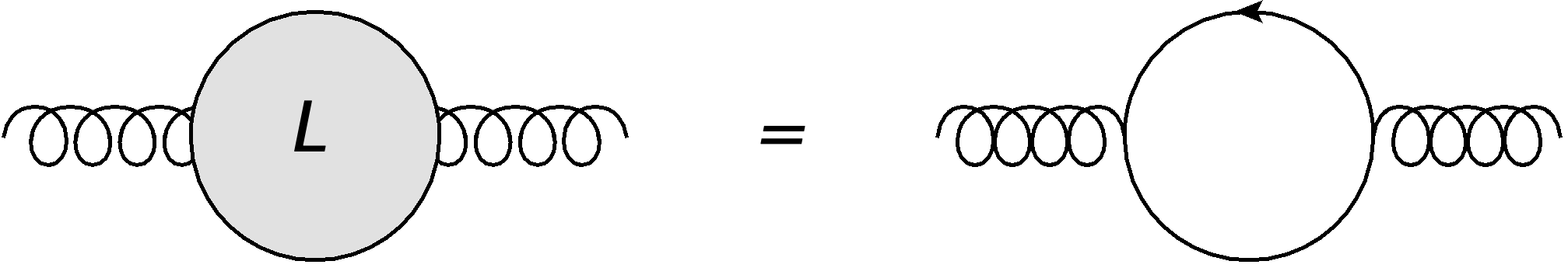}}
\noindent
The scalar form factor to this bare one-loop contribution, as defined in eq.~\eqref{eq:ggTen}, reads
\begin{equation}
i \, \Pi_{L}
= \int \frac{d^d k}{(2\pi)^d} \frac{\mathcal{N}_L(k,p,\ep)}{(-k^2)(-(k+p)^2)} \,.
\end{equation}
where $\mathcal{N}_L$ is the numerator that arises from acting on this Feynman diagram with the projection operator defined in eq.~\eqref{eq:proj}.
After the IBP reduction of all scalar Feynman integrals appearing in this 1PI form factor, we obtain
\begin{equation}
i \, \Pi_{L}
\overset{\text{IBP}}{=\joinrel=} \lambda \, p^2 c_L(\ep) M^{(L)}(p^2,\ep) \,,
\end{equation}
where the only one scalar MI reads
\\
\centerline{\includegraphics[width=0.15\textwidth]{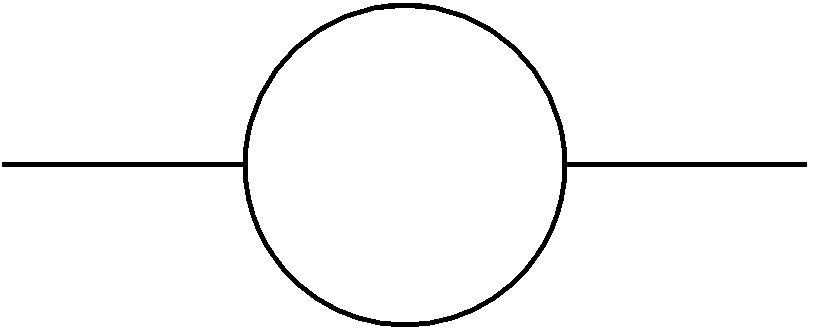}}
\begin{equation}
\begin{split}
    &= M^{(L)}(p^2) = I_{1,1}^{(L)} \\
    &= \int \frac{d^d k}{(2\pi)^d} \frac{1}{(-k^2)(-(k+p)^2)} \\
    &= \frac{i}{(4\pi)^{2-\ep}} \frac{\Gamma(\ep)\Gamma(1-\ep)^2}{\Gamma(2-2\ep)} (-p^2)^{-\ep} \\
    &= \frac{i \, e^{-\ep \gamma_E}}{(4\pi)^{2-\ep}} \left( \frac{1}{\ep} + 2 - \log(-p^2) \right) + \mathcal{O}(\ep) \,,
\end{split}
\label{eq:MIb11}
\end{equation}
in the integral topology defined by a set of generalized propagators
\begin{equation}
\{ \mathcal{D}_i^{(L)} \} = \{ -k^2 , -(k+p)^2 \} \,,
\end{equation}
with a MI coefficient
\begin{equation}
c_{L}(\ep) = \frac{2\ep-2}{3-2\ep} \,.
\end{equation}
By combining the expression for the MI with its coefficient, we obtain
\begin{equation}
\begin{split}    
i \, \Pi_{L}
&= i \, \lambda \, b(\ep) \, p^2 (-p^2)^{-\ep} \\
&= i \, \lambda \, p^2 \frac{e^{-\ep \gamma_E}}{(4\pi)^{2-\ep}}
\left( -\frac{2}{3\ep} -\frac{10}{9} + \frac{2}{3} \log\left(-p^2\right) 
\right) + \mathcal{O}(\ep) \,,
\end{split}
\end{equation}
where we define for convenience an overall factor
\begin{equation}
\begin{split}
b(\ep) 
&= \frac{2\ep-2}{3-2\ep} \frac{1}{(4\pi)^{2-\ep}} \frac{\Gamma(\ep)\Gamma(1-\ep)^2}{\Gamma(2-2\ep)} \\
&= \frac{1}{(4\pi)^{2}} \frac{-2}{3\ep} + \mathcal{O}(\ep^0) \,.
\end{split}
\end{equation}

Note that the resulting bare 1PI gluon propagator $\Pi_L$ needs to be renormalized before geometrically summed into the full propagator $D$.
Indeed, in order to have a meaningful power series, we have to avoid expressions of the form $\lim_{\ep \to 0} \sum_{l=0}^\infty c_l \, \ep^{-l\ep}$.
Each of the $\ep^{-1}$ poles results from the UV singularity of the one-loop two-point function.
For example, in the $\overline{\text{MS}}$ renormalization scheme
\begin{equation}
\lambda 
= \lambda(\mu) \, (4 \pi e^{-\gamma_E})^{-\ep} \, \mu^{2\ep}
\sum_{k=0}^\infty \left( \frac{2 \lambda(\mu)}{3 \ep (4\pi)^2} \right)^k \,,
\end{equation}
the renormalized 1PI gluon propagator $\Pi_{L,\text{ren}}$ at scale $\mu$ reads
\begin{equation}
\Pi_{L,\text{ren}}
= \frac{\lambda(\mu)}{(4\pi)^2} p^2 
\left( -\frac{10}{9} + \frac{2}{3} \log\left(\frac{-p^2}{\mu^2}\right) 
\right) \,,
\end{equation}
while the corresponding full gluon propagator~\footnote{Note also that the geometric summation of the 1PI contributions into the full gluon propagator formally makes sense only for $|\Pi_L(p^2)/p^2| < 1$.
However, following a standard practice in QFT, we are going to treat the resulting function as working for all values of $p^2$, by the virtue of analytic continuation.
This is similar e.g. to the result of Feynman integration converging only for dimensional regulator $\ep$ in a specific region being used for all values of $\ep$.} yields
\begin{equation}
    D^{\mu\nu ab}_L
    = \delta^{ab} \frac{P^{\mu\nu}}{p^2-\Pi_L(p^2)} \,.
\label{eq:Lgeom}
\end{equation}

\subsection{Subleading order}

At subleading order in large $N_f$, the all-loop 1PI gluon propagator 
\\
\centerline{\includegraphics[width=0.8\textwidth]{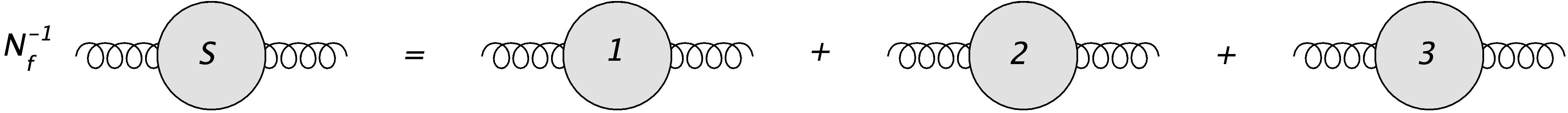}}
\noindent
receives contributions from one-loop-based 
\\
\centerline{\includegraphics[width=0.8\textwidth]{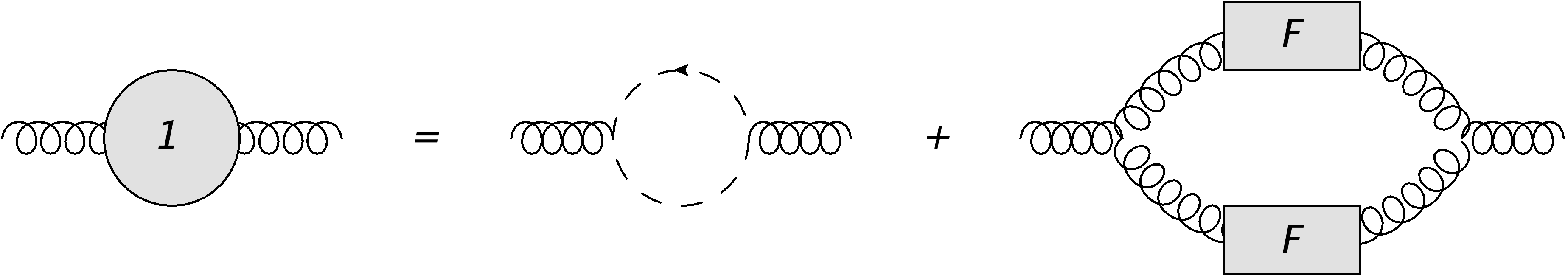}}
\noindent
two-loop-based 
\\
\centerline{\includegraphics[width=0.99\textwidth]{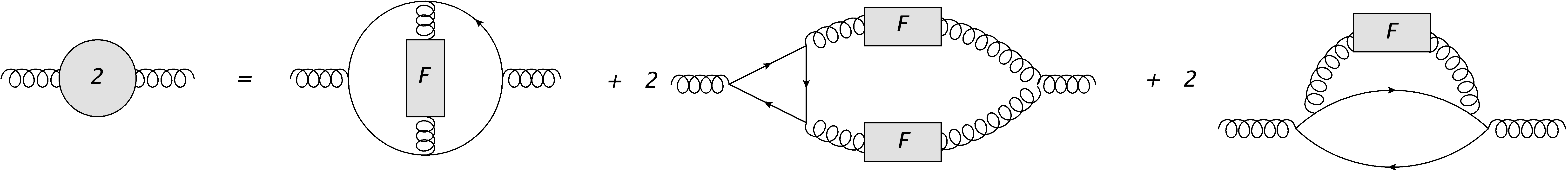}}
\noindent
and three-loop-based 
\\
\centerline{\includegraphics[width=0.9\textwidth]{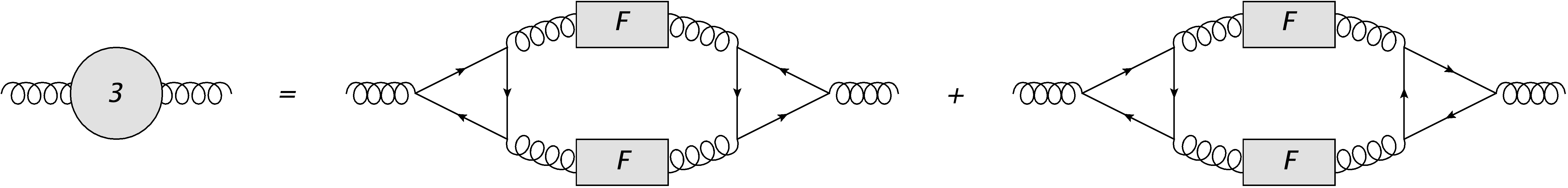}}
\noindent
Feynman diagrams that involve up to infinite quark loop chains
\\
\centerline{\includegraphics[width=0.8\textwidth]{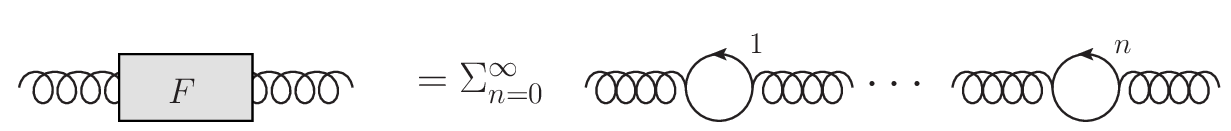}}
\begin{equation}
= \delta^{ab} P^{\mu\nu} \sum_{n=0}^\infty 
\left( \frac{i \, \Pi_L(p^2)}{i \, p^2} \right)^n 
= \delta^{ab} P^{\mu\nu} \sum_{n=0}^\infty 
\left( \lambda \, b(\ep) (-p^2)^{-\ep} \right)^n \,,
\end{equation}
where the sum over $N_f$ quark flavors in each closed quark loop is implicit.
Note that we use a slight abuse of notation since the quark loop chains are summed over to infinity only after the loop momentum integration, and not at the integrand level.
We denote by $N_f^{-1} i \, \Pi_{S,l,n}$ the bare $l$-loop-based subleading contribution arising from the corresponding $n^{\text{th}}$ diagram above.
Let us analyse these contributions case by case, as they correspond to different sets of symbolic powers in integral topologies and have different asymptotics in the large number of quark loops.

\subsubsection{One-loop-based}

Consider the one-loop-based contributions to the all-loop 1PI gluon propagator at subleading order in large $N_f$.

\paragraph{Ghost diagram}\mbox{}\\
The ghost diagram and its result expanded to finite part in $\ep$ reads

\centerline{\includegraphics[width=0.21\textwidth]{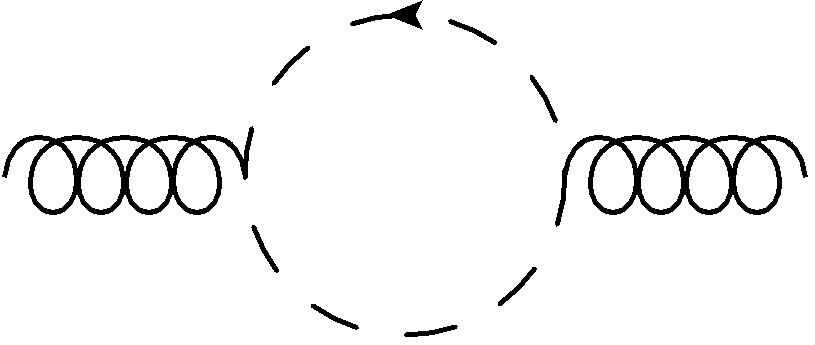}}
\begin{equation}
\begin{split}
= N_f^{-1} i \, \Pi_{S,1,1}(p^2)
&= \int \frac{d^d k}{(2\pi)^d} \frac{\mathcal{N}_{S,1,1}(k,p,\ep)}{(-k^2)(-(k+p)^2)} \\
&\overset{\text{IBP}}{=\joinrel=} \lambda \frac{C_A}{N_f} p^2 c_{S,1,1}(\ep) M^{(S,1,1)}(p^2,\ep) \\
&= \lambda \frac{C_A}{N_f} p^2 \frac{i \, e^{-\ep \gamma_E}}{(4\pi)^{2-\ep}} (-p^2)^{-\ep}
\left( \frac{1}{12\ep} +\frac{2}{9} + \mathcal{O}(\ep) \right) \,,
\end{split}
\end{equation}
where the one-loop bubble MI is the same as at the leading order, see eq.~\eqref{eq:MIb11}, 
\begin{equation}
M^{(S,1,1)} = I_{1,1}^{(L)} = M^{(L)} \,,
\end{equation}
and the MI coefficient is
\begin{equation}
c_{S,1,1}(\ep) = \frac{1}{4(3-2\ep)} \,.
\end{equation}

\paragraph{Two-chain diagram}\mbox{}\\
The two-chain diagram reads

\centerline{\includegraphics[width=0.25\textwidth]{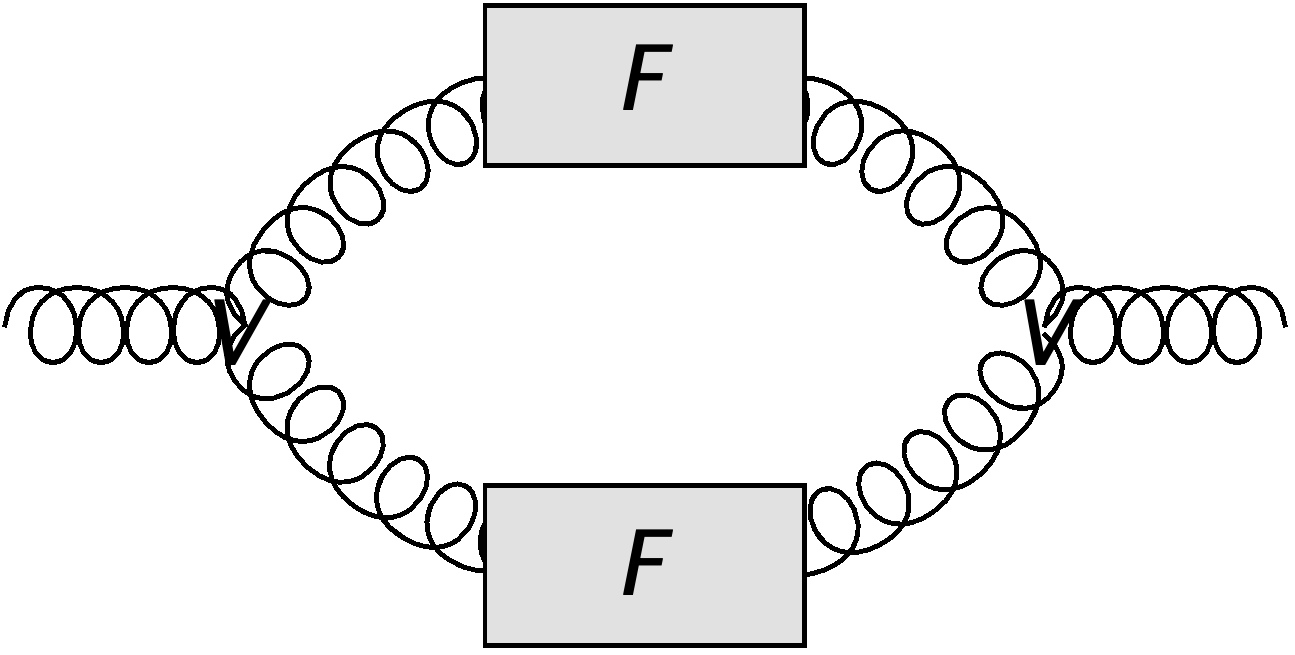}}
\begin{equation}
\begin{split}
&= N_f^{-1} i \, \Pi_{S,1,2}(p^2) 
= \lambda \frac{C_A}{N_f} \sum_{n,m=0}^\infty \int \frac{d^d k}{(2\pi)^d} 
\frac{\mathcal{N}_{S,1,2}(k,p,\ep)}{(-k^2)(-(k+p)^2)} 
\left( \frac{i \, \Pi_{L}(k^2)}{i \, k^2} \right)^n 
\left( \frac{i \, \Pi_{L}((k+p)^2)}{i \, (k+p)^2} \right)^m  \\
&= \lambda \frac{C_A}{N_f} \sum_{n,m=0}^\infty \int \frac{d^d k}{(2\pi)^d} 
\frac{(\lambda \, b(\ep))^{n+m} 
\mathcal{N}_{S,1,2}(k,p,\ep)}{(-k^2)^{1+n\ep}
(-(k+p)^2)^{1+m\ep}} \\
&\overset{\text{IBP}}{=\joinrel=} \lambda \frac{C_A}{N_f} p^2 \sum_{n,m=0}^\infty 
(\lambda \, b(\ep))^{n+m} 
c_{S,1,2}(\ep,n,m) 
M^{(S,1,2)}(p^2,\ep,n,m) \,,
\end{split}
\end{equation}
with the MI coefficient
\begin{equation}
c_{S,1,2}(\ep,n,m) = \frac{25}{12} + \mathcal{O}(\ep)
\end{equation}
and the MI

\centerline{\includegraphics[width=0.2\textwidth]{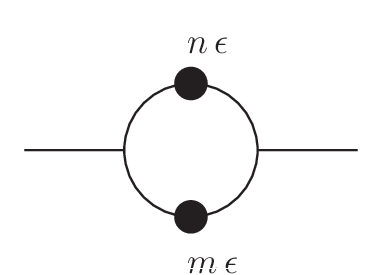}}
\begin{equation}
\begin{split}
&= M^{(S,1,2)}(p^2) = I_{1+n\ep,1+m\ep}^{(L)} \\
&= \int \frac{d^d k}{(2\pi)^d} \frac{1}{(-k^2)^{1+n\ep}(-(k+p)^2)^{1+m\ep}} \\
&= \frac{i}{(4\pi)^{2-\ep}} 
\frac{\Gamma(1-(1+n)\ep)\Gamma(1-(1+m)\ep)\Gamma((1+n+m)\ep)}
{\Gamma(1+n\ep)\Gamma(1+m\ep)\Gamma(2-(2+n+m)\ep)} 
(-p^2)^{-\ep(1+n+m)} \\
&= \frac{i}{(4\pi)^{2}} \frac{1}{\ep \, (1+n+m)} + \mathcal{O}(\ep^0) \,,
\end{split}
\end{equation}
where the label beside the dot in the integral diagram denotes the power of the corresponding propagator minus one, while a lack of a label denotes power two.
The result expanded to finite part in $\ep$ is
\begin{equation}
\begin{split}
N_f^{-1} i \, \Pi_{S,1,2}(p^2) 
&= \lambda \frac{C_A}{N_f} p^2 \sum_{n,m=0}^\infty 
(\lambda \, b(\ep))^{n+m} \frac{i \, e^{-\ep \gamma_E}}{(4\pi)^{2-\ep}} (-p^2)^{-\ep(1+n+m)} \\
&\times \left( \frac{25}{12\ep(1+n+m)} + \frac{47}{72} + \frac{131}{72(1+n+m)} + \mathcal{O}(\ep) \right) \,.
\end{split}
\end{equation}
After simplifying the double sum over $n$ and $m$ into a single sum over all loops $l$, we obtain
\begin{equation}
\Pi_{S,1,2}
= p^2 \frac{\lambda e^{-\ep \gamma_E} (-p^2)^{-\ep}}{(4\pi)^{2-\ep}} \sum_{l=0}^\infty 
(\lambda \, b(\ep) (-p^2)^{-\ep})^l \, \Pi_{S,1,2,l} \,,
\end{equation}
where the $l$-loop contribution has the following large $l$ asymptotics
\begin{equation}
\Pi_{S,1,2,l}
= C_A \left( \frac{25}{12 \epsilon } + \frac{47}{72}l + \frac{89}{36}  + \mathcal{O}(\ep)\right)
\overset{l \to \infty}{\longrightarrow} C_A \left( \frac{25}{12 \epsilon } + \frac{47}{72}l  + \mathcal{O}(\ep) \right) \,.
\end{equation}

\subsubsection{Two-loop-based}

Consider the two-loop-based contributions to the all-loop 1PI gluon propagator at subleading order in large $N_f$.

\paragraph{Sunset diagram}\mbox{}\\
The sunset diagram reads

\centerline{\includegraphics[width=0.21\textwidth]{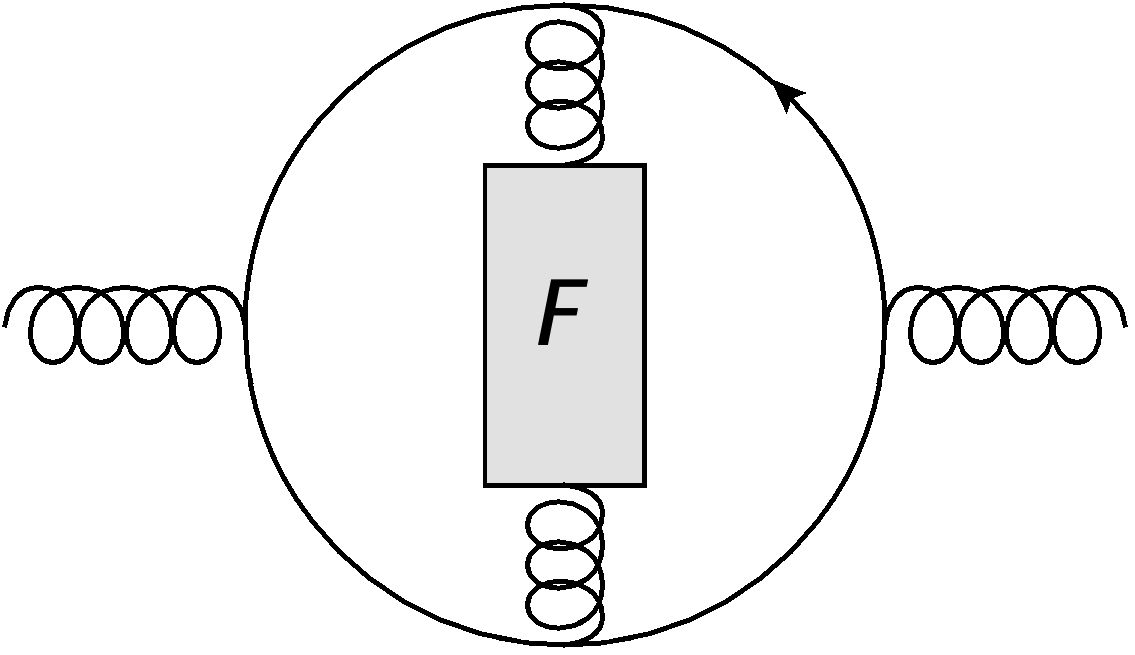}}
\begin{equation}
\begin{split}
&= N_f^{-1} i \, \Pi_{S,2,1}(p^2) = \\
&= \lambda^2 \frac{1}{N_f N_c} \sum_{n=0}^\infty \int \frac{d^d k_1}{(2\pi)^d} \frac{d^d k_2}{(2\pi)^d} 
\frac{\mathcal{N}_{S,2,1}(k_1,k_2,p,\ep)}{(-k_1^2) (-(k_1+p)^2) (-k_2^2) (-(k_2+p)^2) (-(k_1-k_2)^2)} \\
&\times \left( \frac{\Pi_{L}((k_1-k_2)^2)}{(k_1-k_2)^2} \right)^n \\
&= \lambda^2 \frac{1}{N_f N_c} \sum_{n=0}^\infty \int \frac{d^d k_1}{(2\pi)^d} \frac{d^d k_2}{(2\pi)^d} 
\frac{(\lambda \, b(\ep))^{n} 
\mathcal{N}_{S,2,1}(k_1,k_2,p,\ep)}{(-k_1^2) (-(k_1+p)^2) (-k_2^2) (-(k_2+p)^2) (-(k_1-k_2)^2)^{1+n\ep}} \\
&\overset{\text{IBP}}{=\joinrel=} \lambda^2 \frac{1}{N_f N_c} \sum_{n=0}^\infty
(\lambda \, b(\ep))^{n} 
\sum_{i=1}^3 c_{S,2,1,i}(\ep,n) 
M_i^{(S,2,1)}(p^2,\ep,n) \,,
\end{split}
\end{equation}
where we define an integral topology
\begin{equation}
\{ \mathcal{D}_i^{(S,2,1)} \} = \{ -k_1^2 , -k_2^2 , -(k_1-k_2)^2 , -(k_1+p)^2 , -(k_2+p)^2\} \,,
\end{equation}
the MI coefficients start their $\ep$ expansion with
\begin{equation}
\begin{split}
c_{S,2,1,1}(\ep,n) &= -i\ep(1+2n) + \mathcal{O}(\ep^2) \,, \\
c_{S,2,1,2}(\ep,n) &= 
i \frac{2 (\epsilon -1) \left(\left(n^2+3 n+2\right) \epsilon ^3-(2 n+3) \epsilon ^2+3 \epsilon -2\right)}{(3-2 \epsilon ) ((n+2) \epsilon -2) ((n+2) \epsilon -1)} 
\\
&= i\frac{2}{3} + \mathcal{O}(\ep) \,,
\end{split}
\end{equation}
and the MIs are

\centerline{\includegraphics[width=0.25\textwidth]{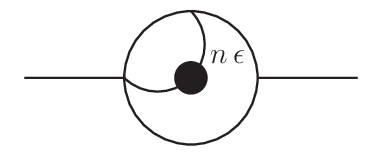}}
\begin{equation}
\begin{split}
&= M_1^{(S,2,1)}
= I_{0,1,1+n\ep,1,1}^{(S,2,1)} \\
&= \left( \frac{i}{(4\pi)^{2-\ep}} \right)^2
\frac{\Gamma (1-\epsilon )^2 \Gamma (-n \epsilon -2 \epsilon +1) \Gamma (-n \epsilon -\epsilon +1) \Gamma (n \epsilon +\epsilon ) \Gamma (n \epsilon +2 \epsilon )}{\Gamma (-n \epsilon -3 \epsilon +2)
   \Gamma (-n \epsilon -2 \epsilon +2) \Gamma (n \epsilon +1) \Gamma (n \epsilon +\epsilon +1)}
(-p^2)^{-\ep(2+n)} \\
&= \left( \frac{i}{(4\pi)^{2}} \right)^2 \frac{1}{\ep^2} \left( \frac{1}{1+n} - \frac{1}{2+n} \right) + \mathcal{O}(\ep^{-1}) \,,
\end{split}
\end{equation}

\centerline{\includegraphics[width=0.25\textwidth]{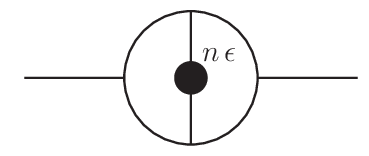}}
\begin{equation}
\begin{split}
&= M_2^{(S,2,1)} 
= p^2 \, I_{1,1,1+n\ep,1,1}^{(S,2,1)} \\
&= \left( \frac{i}{(4\pi)^{2-\ep}} \right)^2
\frac{\Gamma(1+\ep(2+n))}{\Gamma(1+n \ep)} (6 \zeta(3)  + \mathcal{O}(\ep))
(-p^2)^{-\ep(2+n)}
\end{split}
\end{equation}
The result is
\begin{equation}
N_f^{-1} i \, \Pi_{S,2,1}(p^2) 
= p^2 \left( \frac{\lambda e^{-\ep \gamma_E} (-p^2)^{-\ep}}{(4\pi)^{2-\ep}} \right)^2 \sum_{n=0}^\infty 
(\lambda \, b(\ep) (-p^2)^{-\ep})^n \, N_f^{-1} i \, \Pi_{S,2,1,n} \,,
\end{equation}
with the $n$-loop contribution
\begin{equation}
\begin{split}
-N_c \, \Pi_{S,2,1,n} = 
\frac{\frac{1}{n+1}-\frac{3}{n+2}}{\epsilon }+\frac{5}{2 (n+1)}-\frac{17}{6 (n+2)}+\frac{1}{3} (12 \zeta(3)-17)
+ \mathcal{O}(\ep) \,,
\end{split}
\end{equation}
which has the following large $n$ asymptotics
\begin{equation}
\Pi_{S,2,1,n}
\overset{n \to \infty}{\longrightarrow} -N_c \left(
-\frac{2}{n \epsilon } +\frac{1}{3} (12 \zeta(3)-17)
+ \mathcal{O}(\ep) \right) \,.
\end{equation}

\paragraph{One-triangle two-chain diagram}\mbox{}\\
The one-triangle two-chain diagram reads

\centerline{\includegraphics[width=0.3\textwidth]{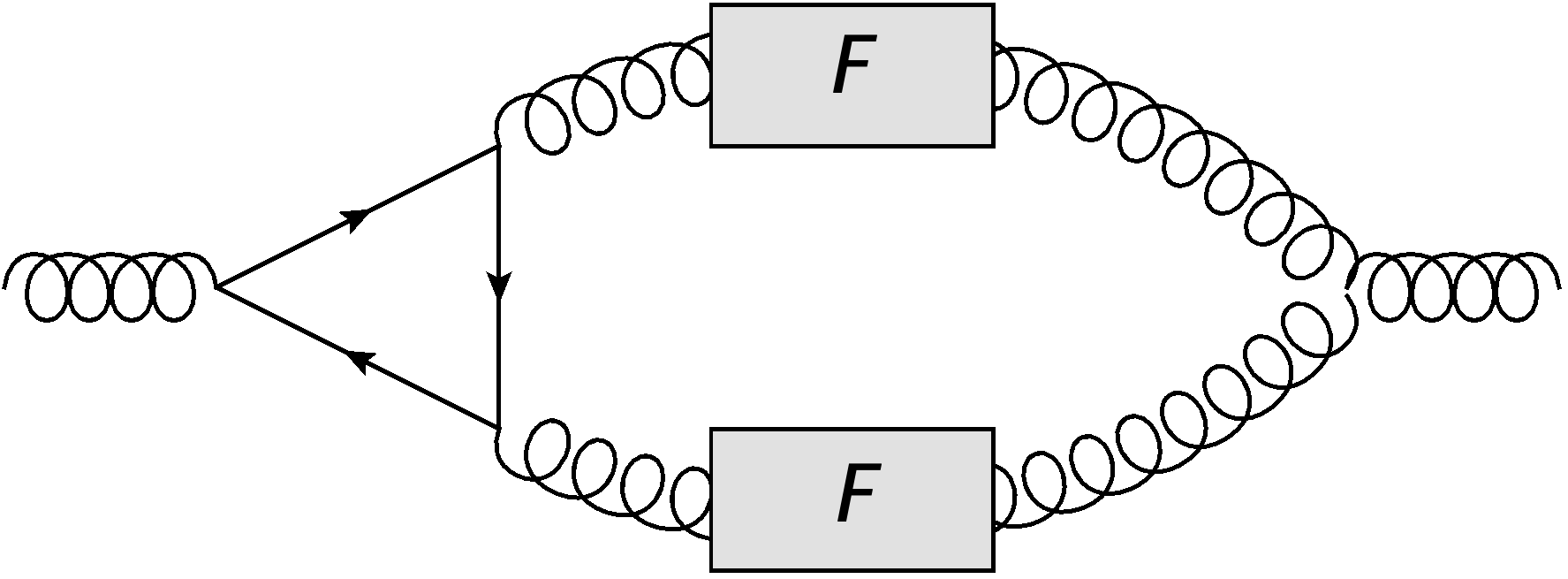}}
\begin{equation}
\begin{split}
&= N_f^{-1} i \, \Pi_{S,2,2}(p^2) = \\
&= \lambda^2 \frac{C_A}{N_f} \sum_{n,m=0}^\infty \int \frac{d^d k_1}{(2\pi)^d} \frac{d^d k_2}{(2\pi)^d} 
\frac{\mathcal{N}_{S,2,2}(k_1,k_2,p,\ep)}{(-k_1^2) (-(k_1+p)^2) (-k_1^2) (-(k_1+p)^2) (-(k_1-k_2)^2)} \\
&\times
\left( \frac{\Pi_{L}(k_2^2)}{k_2^2} \right)^n 
\left( \frac{\Pi_{L}((k_2+p)^2)}{(k_2+p)^2} \right)^m \\
&= \lambda^2 \frac{C_A}{N_f} \sum_{n,m=0}^\infty \int \frac{d^d k_1}{(2\pi)^d} \frac{d^d k_2}{(2\pi)^d} 
\frac{(\lambda \, b(\ep))^{n+m} 
\mathcal{N}_{S,2,2}(k_1,k_2,p,\ep)}{(-k_1^2) (-(k_1+p)^2) (-k_2^2)^{1+n\ep} (-(k_2+p)^2)^{1+m\ep} (-(k_1-k_2)^2)} \\
&\overset{\text{IBP}}{=\joinrel=} \lambda^2 \frac{C_A}{N_f} p^2 \sum_{n,m=0}^\infty 
(\lambda \, b(\ep))^{n+m} 
\sum_{i=1}^3 c_{S,2,2,i}(\ep,n,m) 
M_i^{(S,2,2)}(p^2,\ep,n,m) \,,
\end{split}
\end{equation}
with MI coefficients
\begin{equation}
\begin{split}
c_{S,2,2,1}(\ep,n,m) &= -i \frac{30m-72n-42}{108(1+n)} + \mathcal{O}(\ep) \,, \\
c_{S,2,2,2}(\ep,n,m) &= 5i \frac{m^2-3-2m-2n-3nm+n^2}{18(1+n)(1+m)} + \mathcal{O}(\ep) \,, \\
c_{S,2,2,3}(\ep,n,m) &= -i \frac{5n-7-12m}{18(1+m)} + \mathcal{O}(\ep) \,,
\end{split}
\end{equation}
and MIs

\centerline{\includegraphics[width=0.25\textwidth]{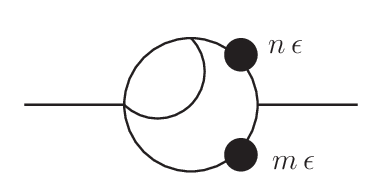}}
\begin{equation}
\begin{split}
&= M_1^{(S,2,2)}
= I_{0,1+n\ep,1,1,1+m\ep}^{(S,2,1)} \\
&= \left( \frac{i}{(4\pi)^{2-\ep}} \right)^2
\frac{\Gamma (1-\epsilon )^2 \Gamma (\epsilon ) \Gamma (-m \epsilon -2 \epsilon +1) \Gamma (-n \epsilon -\epsilon +1) \Gamma (m \epsilon +n \epsilon +2 \epsilon )}{\Gamma (2-2 \epsilon ) \Gamma (m
   \epsilon +\epsilon +1) \Gamma (n \epsilon +1) \Gamma (-m \epsilon -n \epsilon -3 \epsilon +2)}
(-p^2)^{-\ep(2+n+m)} \\
&= \left( \frac{i}{(4\pi)^{2}} \right)^2 \frac{1}{\ep^2(2+m+n)} + \mathcal{O}(\ep^{-1}) \,,
\end{split}
\end{equation}

\centerline{\includegraphics[width=0.3\textwidth]{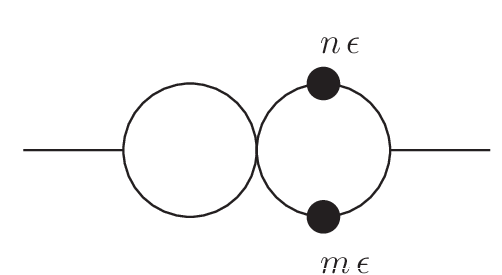}}
\begin{equation}
\begin{split}
&= M_2^{(S,2,2)} 
= I_{1,1+n\ep,0,1,1+m\ep}^{(S,2,1)} 
= I_{1,1}^{(L)} \, I_{1+n\ep,1+m\ep}^{(L)} 
= M^{(L)} \, M^{(S,1,2)}\\
&= \left( \frac{i}{(4\pi)^{2-\ep}} \right)^2
\frac{\Gamma (1-\epsilon )^2 \Gamma (\epsilon ) \Gamma (-m \epsilon -\epsilon +1) \Gamma (-n \epsilon -\epsilon +1) \Gamma (m \epsilon +n \epsilon +\epsilon )}{\Gamma (2-2 \epsilon ) \Gamma (m \epsilon
   +1) \Gamma (n \epsilon +1) \Gamma (-m \epsilon -n \epsilon -2 \epsilon +2)}
(-p^2)^{-\ep(2+n+m)} \\
&= \left( \frac{i}{(4\pi)^{2}} \right)^2 \frac{1}{\ep^2(1+m+n)} + \mathcal{O}(\ep^{-1}) \,,
\end{split}
\end{equation}
\begin{equation}
M_3^{(S,2,2)}
= I_{1,1+n\ep,1,0,1+m\ep}^{(S,2,1)}
= M_1^{(S,2,2)}|_{n \leftrightarrow m}
\end{equation}
The result is
\begin{equation}
N_f^{-1} i \, \Pi_{S,2,2}(p^2) 
= p^2 \left( \frac{\lambda e^{-\ep \gamma_E} (-p^2)^{-\ep}} {(4\pi)^{2-\ep}} \right)^2 \sum_{n,m=0}^\infty 
(\lambda \, b(\ep) (-p^2)^{-\ep})^{n+m} N_f^{-1} i \, \Pi_{S,2,2,n,m} \,,
\end{equation}
with the summands
\begin{align}
-C_A^{-1} \Pi_{S,2,2,n,m}
&= \frac{\frac{17}{9 (m+n+2)}-\frac{25}{18 (m+n+1)}}{\epsilon ^2}
+ \frac{\frac{59}{27 (m+n+2)}-\frac{9}{4 (m+n+1)}+\frac{2}{3}}{\epsilon   } \\ \notag
&+ \frac{150 \pi ^2-2761}{648 (m+n+1)}+\frac{1}{36} (44 m+95)+\frac{13 n}{18} \\ \notag
&+ \frac{-81 m^2-162 m-51 \pi ^2+631}{162 (m+n+2)}
+ \mathcal{O}(\ep) \,.
\end{align}
After simplifying the double sum over $n$ and $m$ into a single sum over all loops $l$, we obtain
\begin{equation}
\Pi_{S,2,2}
= p^2 \left( \frac{\lambda e^{-\ep \gamma_E} (-p^2)^{-\ep}} {(4\pi)^{2-\ep}} \right)^2 \sum_{l=0}^\infty 
(\lambda \, b(\ep) (-p^2)^{-\ep})^{l} \, \Pi_{S,2,2,l} \,,
\end{equation}
where the $l$-loop contribution has the following large $l$ asymptotics
\begin{equation}
\Pi_{S,2,2,l}
\overset{l \to \infty}{\longrightarrow} -C_A \left(
\frac{1}{2 \epsilon ^2}
+\frac{2}{3 \epsilon } l
+ \frac{29}{36} l^2
+ \mathcal{O}(\ep) 
\right) \,.
\end{equation}

\paragraph{Quark propagator correction diagram}\mbox{}\\
The quark propagator correction diagram reads

\centerline{\includegraphics[width=0.25\textwidth]{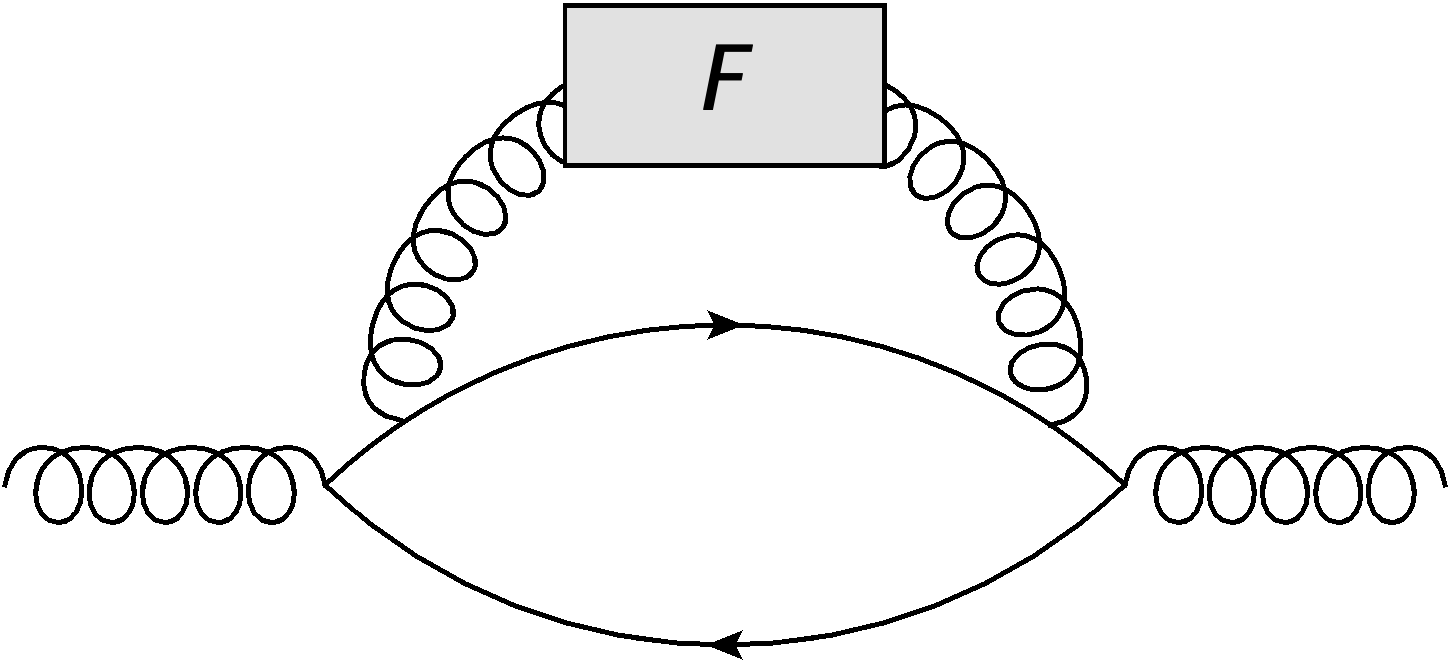}}
\begin{equation}
\begin{split}
&= N_f^{-1} i \, \Pi_{S,2,3}(p^2) = \\
&= \lambda^2 \frac{C_F}{N_f} \sum_{n=0}^\infty \int \frac{d^d k_1}{(2\pi)^d} \frac{d^d k_2}{(2\pi)^d} 
\frac{\mathcal{N}_{S,2,3}(k_1,k_2,p,\ep)}{(-(k_1+p)^2)(-(k_1-k_2)^2) (-k_2^2) (-(k_2+p)^2)^2} 
\left( \frac{\Pi_{L}((k_1+p)^2)}{(k_1+p)^2} \right)^n \\
&= \lambda^2 \frac{C_F}{N_f} \sum_{n=0}^\infty \int \frac{d^d k_1}{(2\pi)^d} \frac{d^d k_2}{(2\pi)^d} 
\frac{(\lambda \, b(\ep))^{n} 
\mathcal{N}_{S,2,3}(k_1,k_2,p,\ep)}{(-(k_1+p)^2)^{1+n\ep}(-(k_1-k_2)^2) (-k_2^2) (-(k_2+p)^2)^2} \\
&\overset{\text{IBP}}{=\joinrel=} \lambda^2 \frac{C_F}{N_f} p^2 \sum_{n=0}^\infty 
(\lambda \, b(\ep))^{n} 
c_{S,2,3}(\ep,n) 
M^{(S,2,3)}(p^2,\ep,n) \,,
\end{split}
\end{equation}
with MI coefficient
\begin{equation}
\begin{split}
c_{S,2,3}(\ep,n)
&= i\frac{2 n (\epsilon -1)^2 \epsilon  (2 \epsilon -3) (2 \epsilon -1)}{(n+1) (n \epsilon +1) ((n+2) \epsilon -2) ((n+2) \epsilon -1) ((n+3) \epsilon -3)} \\
&= -i\ep \frac{n}{1+n} + \mathcal{O}(\ep^2) \,,
\end{split}
\end{equation}
and MI
\begin{equation}
M^{(S,2,3)}
= I_{0,1+n\ep,1,1,1}^{(S,2,1)} 
= M_1^{(S,2,2)}|_{m = 0} \,.
\end{equation}
The result is
\begin{equation}
N_f^{-1} i \, \Pi_{S,2,3}(p^2)
= p^2 \left( \frac{\lambda e^{-\ep \gamma_E} (-p^2)^{-\ep}} {(4\pi)^{2-\ep}} \right)^2 \sum_{n=0}^\infty 
(\lambda \, b(\ep) (-p^2)^{-\ep})^{n} N_f^{-1} i \, \Pi_{S,2,3,n} \,,
\end{equation}
with the $n$-loop contribution
\begin{align}
-C_F^{-1} \Pi_{S,2,3,n}
&= \frac{\frac{1}{n+1}-\frac{2}{n+2}}{\epsilon }+\frac{5}{2 (n+1)}-\frac{4}{3 (n+2)}-\frac{11}{6}
+ \mathcal{O}(\ep) \,,
\end{align}
which has the following large $n$ asymptotics
\begin{equation}
\Pi_{S,2,3,n}
\overset{n \to \infty}{\longrightarrow} -C_F \left(
-\frac{1}{n\ep}-\frac{11}{6}
+ \mathcal{O}(\ep) 
\right) \,.
\end{equation}

\subsubsection{Three-loop-based}
The three-loop-based diagrams read

\centerline{\includegraphics[width=0.7\textwidth]{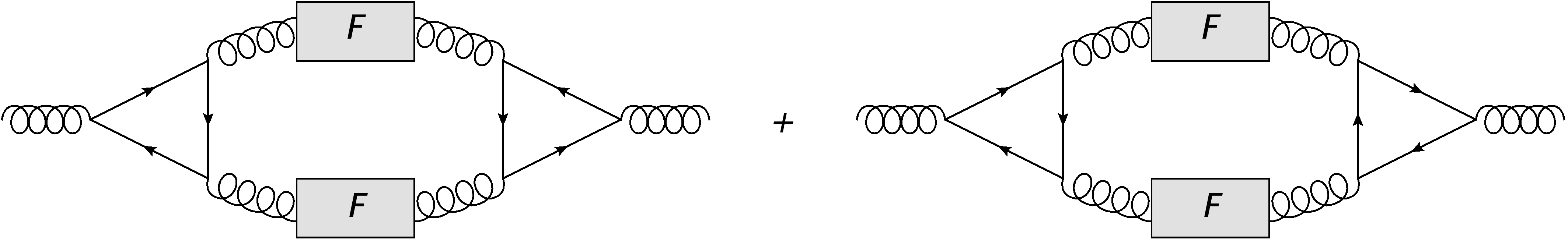}}
\begin{equation}
\begin{split}
&= N_f^{-1} i \, \Pi_{S,3}(p^2) = \\
&= \lambda^3 \frac{\mathcal{C}}{N_f} p^2 \sum_{n,m=0}^\infty 
(\lambda \, b(\ep))^{n+m} 
\sum_{j} c'_{S,3,j}(k,p,\ep) 
I_j^{(S,3)}(p^2,\ep,n,m) \\
&\overset{\text{IBP}}{=\joinrel=} \lambda^3 \frac{\mathcal{C}}{N_f} p^2 \sum_{n,m=0}^\infty 
(\lambda \, b(\ep))^{n+m} 
\sum_{i=1}^7 c_{S,3,i}(\ep,n,m) 
M_i^{(S,3)}(p^2,\ep,n,m) \,,
\end{split}
\end{equation}
where we no longer write the explicit propagator form for the sake of clarity.
The overall color factor receives contributions from the two diagrams (in the two parentheses, respectively)
\begin{equation}
\mathcal{C} 
= \left( \frac{1}{N_c} \right) 
+ \left( C_F - \frac{1}{2N_c} \right)
= \frac{C_A}{2} \,.
\end{equation}
Note that only the second diagram contributes to leading color.
We define the inetgral topology
\begin{equation}
\begin{split}
\{ \mathcal{D}_i^{(S,3)} \} 
=& \{ -k_1^2 ,  -k_2^2 ,  -k_3^2 , -(k_1-k_2)^2 , -(k_2-k_3)^2 , \\
&-(k_3-k_1)^2 , -(k_1+p)^2 , -(k_2+p)^2 , -(k_3+p)^2 \} \,.
\end{split}
\end{equation}
The MI coefficients are
\begin{equation}
\begin{split}
c_{S,3,1}(\ep,n,m) &= 
-\frac{8 (m+2)}{9 (n+1) \epsilon  (m+n+3)} 
+ \mathcal{O}(\ep^0) \,, \\
c_{S,3,2}(\ep,n,m) &= 
\frac{16}{9 (n+1) \epsilon }
+ \mathcal{O}(\ep^0) \,, \\
c_{S,3,3}(\ep,n,m) &= 
-\frac{16}{9 \epsilon  (m+n+3)}
+ \mathcal{O}(\ep^0) \,, \\
c_{S,3,4}(\ep,n,m) &= 
-\frac{8 (m+n+1) (m+n+4)}{9 (m+1) (n+1) \epsilon  (m+n+3)}
+ \mathcal{O}(\ep^0) \,, \\
c_{S,3,5}(\ep,n,m) &= 
\frac{16}{9 (m+1) \epsilon }
+ \mathcal{O}(\ep^0) \,, \\
c_{S,3,6}(\ep,n,m) &= 
-\frac{8 (n+2)}{9 (m+1) \epsilon  (m+n+3)}
+ \mathcal{O}(\ep^0) \,, \\
c_{S,3,7}(\ep,n,m) &= \mathcal{O}(\ep) \,,
\end{split}
\end{equation}
and the MIs are

\centerline{\includegraphics[width=0.2\textwidth]{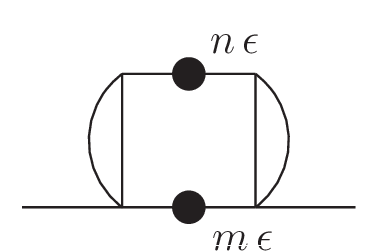}}
\begin{equation}
\begin{split}
&= M_1^{(S,3)}
= I_{0,1+n\ep,0,1,1,0,1,1+m\ep,1}^{(S,3)} \\
&= \left( \frac{i}{(4\pi)^{2-\ep}} \right)^3
\frac{\Gamma (1-\epsilon )^4 \Gamma (\epsilon )^2 \Gamma (1-(m+3) \epsilon ) \Gamma (1-(n+1) \epsilon ) \Gamma ((m+n+3) \epsilon )}{\Gamma (2-2 \epsilon )^2 \Gamma ((m+2) \epsilon +1) \Gamma (n \epsilon +1) \Gamma (2-(m+n+4) \epsilon )}
(-p^2)^{-\ep(3+n+m)} \\
&= \left( \frac{i}{(4\pi)^{2}} \right)^3 \frac{1}{\ep^3(3+m+n)} + \mathcal{O}(\ep^{-2}) \,,
\end{split}
\end{equation}

\centerline{\includegraphics[width=0.3\textwidth]{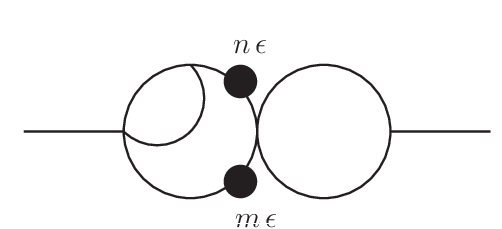}}
\begin{equation}
= M_2^{(S,3)}
= I_{0,1+n\ep,1,1,0,0,1,1+m\ep,1}^{(S,3)} 
= I_{0,1+n\ep,1,1,1+m\ep}^{(S,2,1)} \, I_{1,1}^{(L)} 
= M_1^{(S,2,2)} \, M^{(L)}
\end{equation}

\centerline{\includegraphics[width=0.2\textwidth]{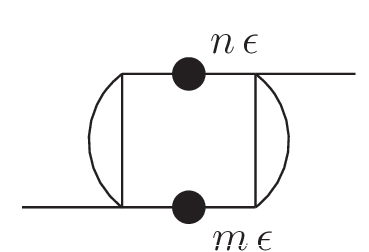}}
\begin{equation}
\begin{split}
&= M_3^{(S,3)}
= I_{0,1+n\ep,1,1,1,0,1,1+m\ep,0}^{(S,3)} \\
&= \left( \frac{i}{(4\pi)^{2-\ep}} \right)^3
\frac{\Gamma (1-\epsilon )^4 \Gamma (\epsilon )^2 \Gamma (1-(m+2) \epsilon ) \Gamma (1-(n+2) \epsilon ) \Gamma ((m+n+3) \epsilon )}{\Gamma (2-2 \epsilon )^2 \Gamma (m \epsilon +\epsilon +1) \Gamma (n \epsilon +\epsilon +1) \Gamma (2-(m+n+4) \epsilon )}
(-p^2)^{-\ep(3+n+m)} \\
&= \left( \frac{i}{(4\pi)^{2}} \right)^3 \frac{1}{\ep^3(3+m+n)} + \mathcal{O}(\ep^{-2}) \,,
\end{split}
\end{equation}

\centerline{\includegraphics[width=0.3\textwidth]{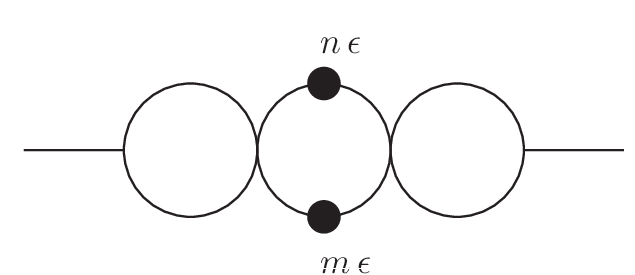}}
\begin{equation}
= M_4^{(S,3)}
= I_{1,1+n\ep,1,0,0,0,1,1+m\ep,1}^{(S,3)} 
= I_{1+n\ep,1+m\ep}^{(L)} \, (I_{1,1}^{(L)})^2
= M^{(S,1,2)} \, (M^{(L)})^2
\end{equation}

\begin{equation}
M_5^{(S,3)}
= I_{1,1+n\ep,1,1,0,0,0,1+m\ep,1}^{(S,3)} 
= M_2^{(S,3)}|_{n \leftrightarrow m}
\end{equation}

\begin{equation}
M_6^{(S,3)}
= I_{1,1+n\ep,1,1,1,0,0,1+m\ep,0}^{(S,3)}
= M_1^{(S,3)}|_{n \leftrightarrow m}
\end{equation}

\centerline{\includegraphics[width=0.3\textwidth]{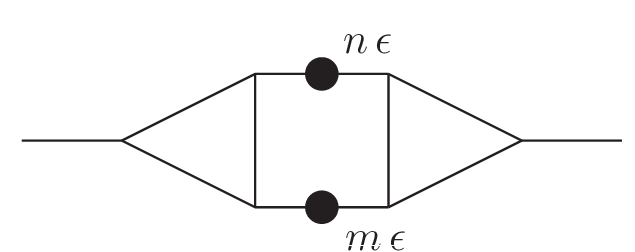}}
\begin{equation}
\begin{split}
&= M_7^{(S,3)}
= (p^2)^2 \, I_{1,1+n\ep,1,1,1,0,0,1+m\ep,0}^{(S,3)} = \mathcal{O}(\ep^{0}) \\
&= \left( \frac{i}{(4\pi)^{2-\ep}} \right)^3
\frac{\Gamma(2+(3+n+m)\ep)}{\Gamma(1+n \ep)\Gamma(1+m \ep)} (20 \zeta(5)  + \mathcal{O}(\ep))
(-p^2)^{-\ep(3+n+m)} \,,
\end{split}
\end{equation}
The result yields
\begin{equation}
N_f^{-1} i \, \Pi_{S,3}(p^2) 
= p^2 \left( \frac{\lambda e^{-\ep \gamma_E} (-p^2)^{-\ep}} {(4\pi)^{2-\ep}} \right)^3 \sum_{n,m=0}^\infty 
(\lambda \, b(\ep) (-p^2)^{-\ep})^{n+m} N_f^{-1} i \, \Pi_{S,3,n,m} \,,
\end{equation}
with the summands
\begin{align}
-\mathcal{C}^{-1} \Pi_{S,3,n,m}
&= \frac{\frac{4}{3 (m+n+2)}-\frac{86}{27 (m+n+3)}}{\epsilon ^3} \\ \notag
&+ \frac{-\frac{505}{81 (m+n+3)}+\frac{38}{9 (m+n+2)}-\frac{47}{81}}{\epsilon ^2} \\ \notag
&+ \frac{\frac{-216 m^2-648 m+387 \pi ^2-7553}{486 (m+n+3)}+\frac{301-9 \pi ^2}{27 (m+n+2)}+\frac{1}{486} (1653-121   m)-\frac{337 n}{486}}{\epsilon } \\ \notag
&- \frac{5 \left(648 m^2+1944 m-9288 \zeta (3)-909 \pi ^2+21367\right)}{2916 (m+n+3)} \\ \notag
&+ \frac{109 m^2-5616 m \zeta (3)+18025 m-60480 \zeta (3)+423 \pi ^2+104129}{2916} \\ \notag
&+ \frac{n (10800 m \zeta (3)-1726 m-5616 \zeta (3)+14785)}{2916} \\ \notag
&+ \frac{-1080 \zeta   (3)+4471-171 \pi ^2}{162 (m+n+2)} + \frac{109 n^2}{2916}
+ \mathcal{O}(\ep) \,.
\end{align}
After simplifying the double sum over $n$ and $m$ into a single sum over all loops $l$, we obtain
\begin{equation}
\Pi_{S,3}
= p^2 \left( \frac{\lambda e^{-\ep \gamma_E} (-p^2)^{-\ep}} {(4\pi)^{2-\ep}} \right)^3 \sum_{l=0}^\infty 
(\lambda \, b(\ep) (-p^2)^{-\ep})^{l} \, \Pi_{S,3,l} \,,
\end{equation}
where the $l$-loop contribution has the following large $l$ asymptotics
\begin{equation}
\Pi_{S,3,l}
\overset{l \to \infty}{\longrightarrow} \frac{-C_A}{2} \left(
-\frac{50}{27 \epsilon ^3}
-\frac{47}{81 \epsilon ^2} l
-\frac{301}{486 \epsilon } l^2 
+ \frac{5 (360 \zeta (3)-43)}{2916} l^3 
+ \mathcal{O}(\ep) 
\right) \,.
\end{equation}

\section{Quark propagator}
\label{sec:qq}

In this section, we calculate the bare full quark propagator $S$ to all loops in the massless subleading large $N_f$ approximation of QCD.

\subsection{General form}

Consider the quark propagator $S$.
We describe a general form of the quark propagator $S$ in a similar way as in case of the gluon propagator $\Pi$ in sec.~\eqref{sec:gg.gen}.
The free quark propagator $S_0$ at momentum $p$ reads
\begin{equation}
\includegraphics[width=0.18\textwidth]{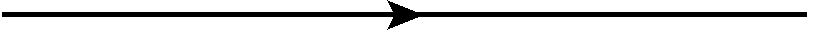} 
= i \, S_{0,jk}(p) 
= \delta_{jk} \, \slashed{p} \, \frac{i}{p^2} \,,
\end{equation}
where $j,k=1,...,N_c$ are the color indices in fundamental representation.
The beyond-tree-level 1PI quark propagator $\Sigma$ has the following tensor form
\begin{equation}
    \includegraphics[width=0.1\textwidth]{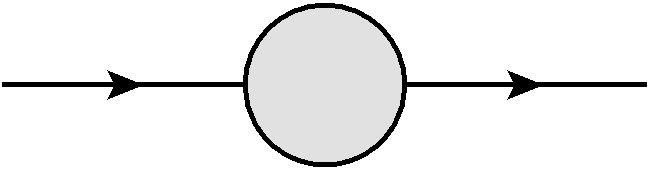}
    = \frac{1}{i} \, \Sigma_{jk}(p) 
    = \delta_{jk} \, \slashed{p} \, \frac{1}{i} \, \Sigma(p^2) \,,
\end{equation}
which is universal for all the corrections considered below.
In massless QCD, note the absence of the term $S_m \, \mathbb{1}$ in eq.~\eqref{eq:introTensors}.
The scalar form factor can be defined by a projection
\begin{equation}
\Sigma(p^2) 
= \frac{\delta_{jk}}{N_c} \frac{\slashed{p}}{4 \, p^2} \Sigma_{jk}(p) \,.
\end{equation}
The full quark propagator $S$ results from summing up all 1PI contributions $\Sigma$ in a geometric series \\
\includegraphics[width=0.8\textwidth]{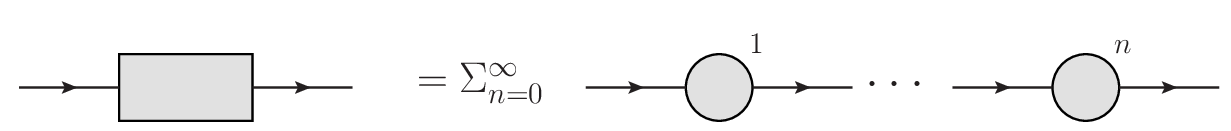}
\begin{equation}
\begin{split}
    i \, S_{jk}(p) 
    &= i \, S_{0,jk}(p) 
    + i \, S_{0,jl}(p) \, \frac{1}{i} \, \Sigma_{ln}(p) \, i \, S_{0,nk}(p) 
    + ... \\
    &= \delta_{jk} \, \slashed{p} \, \frac{i}{p^2 - \Sigma(p^2)} \,.
\end{split}
\end{equation}
In the large $N_f$ approximation, we split the 1PI quark propagator $\Sigma$ into leading $(L)$ and subleading $(S)$ terms
\begin{equation}
    \Sigma = \Sigma_L + N_f^{-1} \, \Sigma_S  + \mathcal{O}(N_f^{-2}) \,.
\end{equation}
One can see from Feynman diagrams that the leading contribution vanishes here
\begin{equation}
    \Sigma_L = 0 \,.
\end{equation}
The corresponding full quark propagator $S$ reads
\begin{equation}
\begin{split}
    S_{jk}(p) 
    &= N_f^{-1} S_{jk,S}(p) + \mathcal{O}(N_f^{-2}) \\
    &= \delta_{jk} \, \slashed{p} \, \frac{i}{p^2 - N_f^{-1} \,\Sigma_S(p^2)} + \mathcal{O}(N_f^{-2})
\end{split}
\end{equation}
to subleading order in large $N_f$.
Let us now discuss the subleading correction.

\subsection{Subleading order}

At subleading order in large $N_f$, the all-loop 1PI quark propagator reads

\centerline{\includegraphics[width=0.5\textwidth]{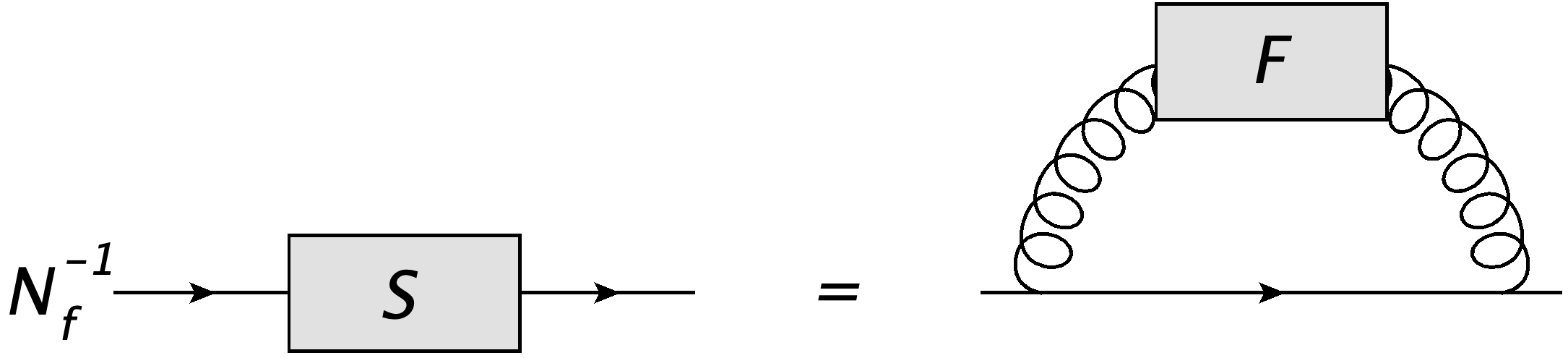}}
\begin{equation}
\begin{split}
&= \frac{1}{i \, N_f} \, \Sigma_{S}(p^2) = \\
&= \lambda \frac{C_F}{N_f} p^2 \sum_{n=0}^\infty 
(\lambda \, b(\ep))^{n} 
\sum_{j} c'(k,p,\ep) 
I_j^{(L)}(p^2,\ep,n) \\
&\overset{\text{IBP}}{=\joinrel=} \lambda \frac{C_F}{N_f} p^2 \sum_{n=0}^\infty 
(\lambda \, b(\ep))^{n} 
c(\ep,n) 
M(p^2,\ep,n) \,,
\end{split}
\end{equation}
with MI coefficient
\begin{equation}
\begin{split}
c(\ep,n)
&= \frac{n \epsilon  \left(2 \epsilon ^2-5 \epsilon +3\right)}{(n \epsilon +1) ((n+2) \epsilon -2)}
\end{split}
\end{equation}
and MI

\centerline{\includegraphics[width=0.2\textwidth]{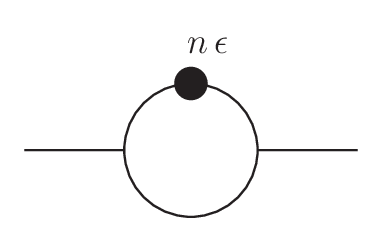}}
\begin{equation}
= M = I_{1+n\ep,1}^{(L)} = M^{(S,1,2)}|_{m = 0}
\end{equation}
The result is
\begin{equation}
\frac{1}{i \, N_f} \, \Sigma_{S}(p^2)
= p^2 \frac{\lambda e^{-\ep \gamma_E} (-p^2)^{-\ep}} {(4\pi)^{2-\ep}}  \sum_{n=0}^\infty 
(\lambda \, b(\ep) (-p^2)^{-\ep})^{n} \frac{1}{i \, N_f} \, \Sigma_{S,n} \,,
\end{equation}
where the summands have the following large $n$ asymptotics
\begin{align}
\Sigma_{S,n}
&= -C_F \left( \frac{3}{2 (n+1)}-\frac{3}{2}
+ \mathcal{O}(\ep) \right)
\overset{n \to \infty}{\longrightarrow} -C_F \left(
-\frac{3}{2}
+ \mathcal{O}(\ep) 
\right) \,.
\end{align}
Contrarily to the gluon propagator, this quark propagator does not have any $\ep^{-1}$ poles, since it is UV finite.
Note that our approximation can be further simplified if $C_F \ll C_A$.
Indeed, if terms proportional to $C_F$ are dropped, the full quark propagator is equal to the free quark propagator, i.e. $S = S_0$.
It is a valid limit since the coefficients of Casimir invariants are separately gauge invariant.
In addition, it also preserves the IR freedom as there is no $C_F$ contribution to the beta function~\eqref{eq:beta}.

\section{Four-quark Bethe-Salpeter kernel}
\label{sec:BSker}

In this section, we calculate the bare four-quark BS kernel $K$ to all loops in the massless subleading large $N_f$ approximation of QCD.

\subsection{General form}
\label{sec:BSker.gen}

Consider the four-quark BS kernel $K$ for the process
\begin{equation}
f(-q_1) + \bar{f'}(-q_2) \to \bar{f'}(k) + f(-k-q_{12})
\end{equation}
labeled as in the BSE~\eqref{eq:BS}.
As momentum $k$ needs to be integrated over in the BSE~\eqref{eq:BS}, two out of the four external massless quarks are off-shell.
Note that it is enough for compute the kernel $K_{f\bar{f'} \to \bar{f'}f}$ in the $f' \neq f$ case in order to reconstruct the $f' = f$ case using crossing symmetry for fermions, i.e.
\begin{equation}
K_{f\bar{f} \to \bar{f}f} 
= K_{f\bar{f} \to \bar{f'}f'} - K_{f\bar{f'} \to \bar{f'}f} \,.
\end{equation}
For convenience, we are going to use new labels $p_i$ to denote the external momenta
\begin{equation}
f(-p_1) + \bar{f}(-p_2) \to f'(p_3) + \bar{f'}(p_4)
\end{equation}
with two on-shell $p_2^2 = 0 = p_3^2$ and two off-shell $p_1^2 \neq 0 \neq p_4^2$ particles.
They are related to the BSE~\eqref{eq:BS} parametrization via
\begin{equation}
p_1 = -k-q_{12} \,, \quad
p_2 = q_1 \,, \quad
p_3 = q_2 \,, \quad
p_4 = k \,.
\end{equation}
The corresponding momentum conservation reads
\begin{equation}
p_1+p_2+p_3+p_4 = 0
\end{equation}
or 
\begin{equation}
s + t + u = p_1^2 + p_4^2
\end{equation}
in Mandelstam variables
\begin{equation}
s = s_{12} \,, \quad
u = s_{23} \,, \quad
t = s_{13} \,,
\end{equation}
with $s_{ij} = (p_i+p_j)^2$.
For further convenience, we also define here $p_{ij} = p_i+p_j$ and $p_{ijk} = p_i+p_j+p_k$.

The four-quark BS kernel $K$ can be linearly decomposed into a basis of color $\mathcal{C}_c$ and Dirac $T_t$ tensors

\centerline{\includegraphics[width=0.15\textwidth]{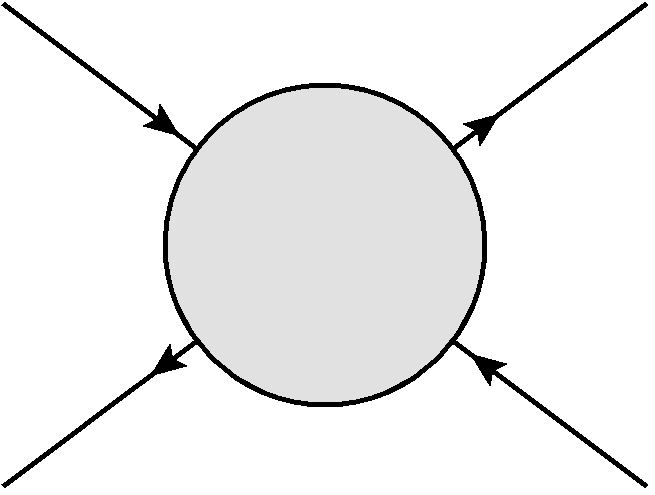}}
\begin{equation}
= K 
= \sum_{t} \, T_t \, K_{t} 
= \sum_{c,t} \mathcal{C}_c \, T_t \, K_{c,t}(s,t,u,p_1^2,\ep) \,.
\end{equation}
The color tensor basis stays the same at each loop order
\begin{equation}
\mathcal{C}_1 = 
\delta_{i_1 i_2} \, \delta_{i_3 i_4} \,, \qquad \mathcal{C}_2 = 
\delta_{i_1 i_4} \, \delta_{i_2 i_3} \,,
\end{equation}
where $i_n=1,...,N_c$ are the four quark color indices in fundamental representation.
In contrary, the Dirac tensor basis $T_t$ depends on the loop order.
Indeed, since we calculate a correlator and not a scattering amplitude, we do not contract our tensors with external Dirac fermion states.
As we need to consider up to one-loop-based corrections, we have 42 such tensors
\begin{equation}
\begin{split}
&T_{1}= (\slashed{p}_1 , \slashed{p}_1) \,, \, \, T_{2}= (\slashed{p}_1 , \slashed{p}_2) \,, \, \, T_{3}= (\slashed{p}_1 , \slashed{p}_3) \,, \, \, \\ &T_{4}= (\slashed{p}_1 , \slashed{p}_1 \slashed{p}_2 \slashed{p}_3) \,, \, \, T_{5}= (\slashed{p}_2 , \slashed{p}_1) \,, \, \, T_{6}= (\slashed{p}_2 , \slashed{p}_2) \,, \, \, \\ &T_{7}= (\slashed{p}_2 , \slashed{p}_3) \,, \, \, T_{8}= (\slashed{p}_2 , \slashed{p}_1 \slashed{p}_2 \slashed{p}_3) \,, \, \, T_{9}= (\slashed{p}_3 , \slashed{p}_1) \,, \, \, \\ &T_{10}= (\slashed{p}_3 , \slashed{p}_2) \,, \, \, T_{11}= (\slashed{p}_3 , \slashed{p}_3) \,, \, \, T_{12}= (\slashed{p}_3 , \slashed{p}_1 \slashed{p}_2 \slashed{p}_3) \,, \, \, \\ &T_{13}= (\slashed{p}_1 \slashed{p}_2 \slashed{p}_3 , \slashed{p}_1) \,, \, \, T_{14}= (\slashed{p}_1 \slashed{p}_2 \slashed{p}_3 , \slashed{p}_2) \,, \, \, T_{15}= (\slashed{p}_1 \slashed{p}_2 \slashed{p}_3 , \slashed{p}_3) \,, \, \, \\ &T_{16}= (\slashed{p}_1 \slashed{p}_2 \slashed{p}_3 , \slashed{p}_1 \slashed{p}_2 \slashed{p}_3) \,, \, \, T_{17}= (\gamma^\mu , \gamma^\mu) \,, \, \, T_{18}= (\gamma^\mu , \gamma^\mu \slashed{p}_1 \slashed{p}_2) \,, \, \, \\ &T_{19}= (\gamma^\mu , \gamma^\mu \slashed{p}_1 \slashed{p}_3) \,, \, \, T_{20}= (\gamma^\mu , \gamma^\mu \slashed{p}_2 \slashed{p}_3) \,, \, \, T_{21}= (\gamma^\mu \slashed{p}_1 \slashed{p}_2 , \gamma^\mu) \,, \, \, \\ &T_{22}= (\gamma^\mu \slashed{p}_1 \slashed{p}_2 , \gamma^\mu \slashed{p}_1 \slashed{p}_2) \,, \, \, T_{23}= (\gamma^\mu \slashed{p}_1 \slashed{p}_2 , \gamma^\mu \slashed{p}_1 \slashed{p}_3) \,, \, \, T_{24}= (\gamma^\mu \slashed{p}_1 \slashed{p}_2 , \gamma^\mu \slashed{p}_2 \slashed{p}_3) \,, \, \, \\ &T_{25}= (\gamma^\mu \slashed{p}_1 \slashed{p}_3 , \gamma^\mu) \,, \, \, T_{26}= (\gamma^\mu \slashed{p}_1 \slashed{p}_3 , \gamma^\mu \slashed{p}_1 \slashed{p}_2) \,, \, \, T_{27}= (\gamma^\mu \slashed{p}_1 \slashed{p}_3 , \gamma^\mu \slashed{p}_1 \slashed{p}_3) \,, \, \, \\ &T_{28}= (\gamma^\mu \slashed{p}_1 \slashed{p}_3 , \gamma^\mu \slashed{p}_2 \slashed{p}_3) \,, \, \, T_{29}= (\gamma^\mu \slashed{p}_2 \slashed{p}_3 , \gamma^\mu) \,, \, \, T_{30}= (\gamma^\mu \slashed{p}_2 \slashed{p}_3 , \gamma^\mu \slashed{p}_1 \slashed{p}_2) \,, \, \, \\ &T_{31}= (\gamma^\mu \slashed{p}_2 \slashed{p}_3 , \gamma^\mu \slashed{p}_1 \slashed{p}_3) \,, \, \, T_{32}= (\gamma^\mu \slashed{p}_2 \slashed{p}_3 , \gamma^\mu \slashed{p}_2 \slashed{p}_3) \,, \, \, T_{33}= (\gamma^\mu \gamma^\nu \slashed{p}_1 , \gamma^\mu \gamma^\nu \slashed{p}_1) \,, \, \, \\ &T_{34}= (\gamma^\mu \gamma^\nu \slashed{p}_1 , \gamma^\mu \gamma^\nu \slashed{p}_2) \,, \, \, T_{35}= (\gamma^\mu \gamma^\nu \slashed{p}_1 , \gamma^\mu \gamma^\nu \slashed{p}_3) \,, \, \, T_{36}= (\gamma^\mu \gamma^\nu \slashed{p}_2 , \gamma^\mu \gamma^\nu \slashed{p}_1) \,, \, \, \\ &T_{37}= (\gamma^\mu \gamma^\nu \slashed{p}_2 , \gamma^\mu \gamma^\nu \slashed{p}_2) \,, \, \, T_{38}= (\gamma^\mu \gamma^\nu \slashed{p}_2 , \gamma^\mu \gamma^\nu \slashed{p}_3) \,, \, \, T_{39}= (\gamma^\mu \gamma^\nu \slashed{p}_3 , \gamma^\mu \gamma^\nu \slashed{p}_1) \,, \, \, \\ &T_{40}= (\gamma^\mu \gamma^\nu \slashed{p}_3 , \gamma^\mu \gamma^\nu \slashed{p}_2) \,, \, \, T_{41}= (\gamma^\mu \gamma^\nu \slashed{p}_3 , \gamma^\mu \gamma^\nu \slashed{p}_3) \,, \, \, T_{42}= (\gamma^\mu \gamma^\nu \gamma^\rho , \gamma^\mu \gamma^\nu \gamma^\rho) \,, \, \,
\end{split}
\end{equation}
where the bracket $(\cdot , \cdot)$ denotes contributions to the two different spinor lines. 
The coefficients $K_t$ of the BS kernel $K$ in this tensor basis $T_i$ can be extracted by acting with a Hermitian conjugate tensor $T_i^\dagger$ on the original kernel
\begin{equation}
K_t = \sum_{i=1}^{42} (A^{-1})_{ti} \sum_{\text{pol}} T_i^\dagger \, K \,,
\end{equation}
while accounting for the inverse of the matrix
\begin{equation}
A_{ti}(s,t,u,p_1^2,\ep) = \sum_{\text{pol}} T_t^\dagger \, T_i \,.
\end{equation}
This matrix with multivariate rational function elements can be inverted using e.g. \texttt{Finite\-Flow}~\cite{Panzer:2014caa}.
We provide it in the ancillary files.
In our further calculation, we focus here on the coefficients
\begin{equation}
F_i = \sum_{\text{pol}} T_i^\dagger \, K \,,
\end{equation}
which are form factors in the corresponding transformed basis.
In the large $N_f$ approximation, we split the form factors $F$ into leading $(L)$ and subleading $(S)$ terms
\begin{equation}
    F = F_L + N_f^{-1} \, F_S  + \mathcal{O}(N_f^{-2}) \,.
\end{equation}

\subsection{Leading order}

At leading order in large $N_f$, the BS kernel is tree-level-based, and for the two quark-antiquark pairs of different flavor $f_1\bar{f}_2$ and $f'_3\bar{f}'_4$, it receives contribution only from one $s$-channel diagram

\centerline{\includegraphics[width=0.5\textwidth]{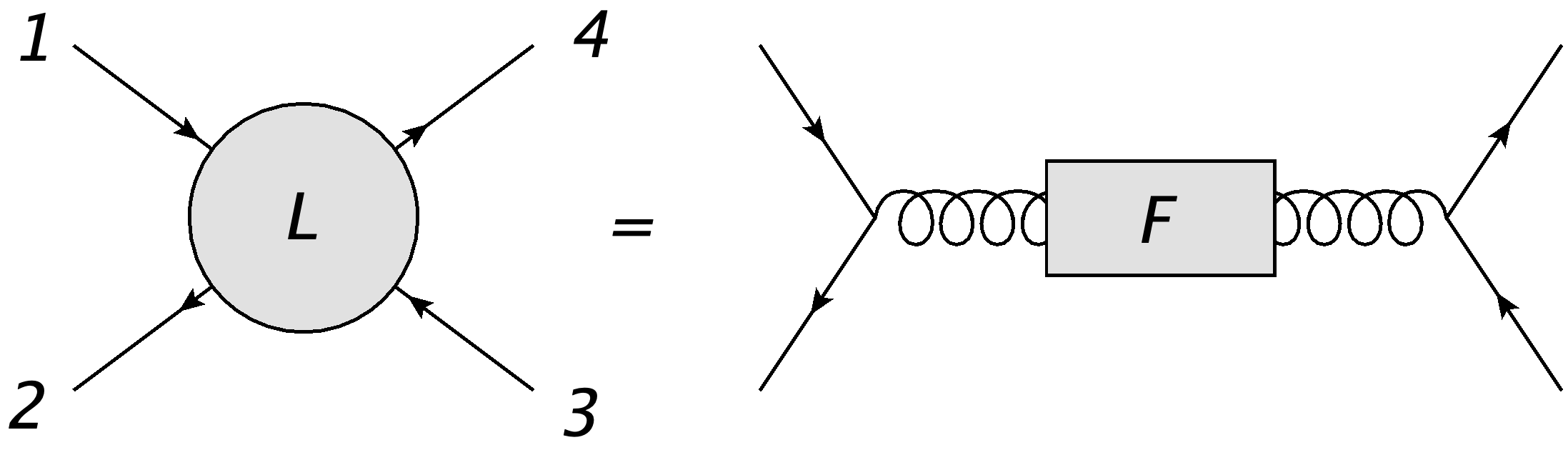}}
\begin{equation}
\begin{split}
&= K_L 
= (i g)^2 \, T^a_{i_1 i_2} T^b_{i_3 i_4} (\gamma_\mu \,, \gamma_\nu) \, \delta^{ab} P^{\mu\nu} \left( \frac{i \, \Pi_{L}(s)}{i \, s} \right)^n  \\
&= i g^2 \left( \frac{\mathcal{C}_1}{N_c} - \mathcal{C}_2 \right) 
\sum_{t,i=1}^{42} (A^{-1})_{ti} \, F_{L,i}(s,t,u,p_1^2,\ep) \, T_t 
\sum_{n=0}^\infty \left( \frac{\Pi_{L}(s)}{s} \right)^n \,,
\end{split}
\end{equation}
where $F_{L,i}$ are attached in the ancillary files.

\subsection{Subleading order}

At subleading order in large $N_f$, the BS kernel receives contributions from a mixture of tree-level-based and one-loop-based corrections
\\
\centerline{\includegraphics[width=0.99\textwidth]{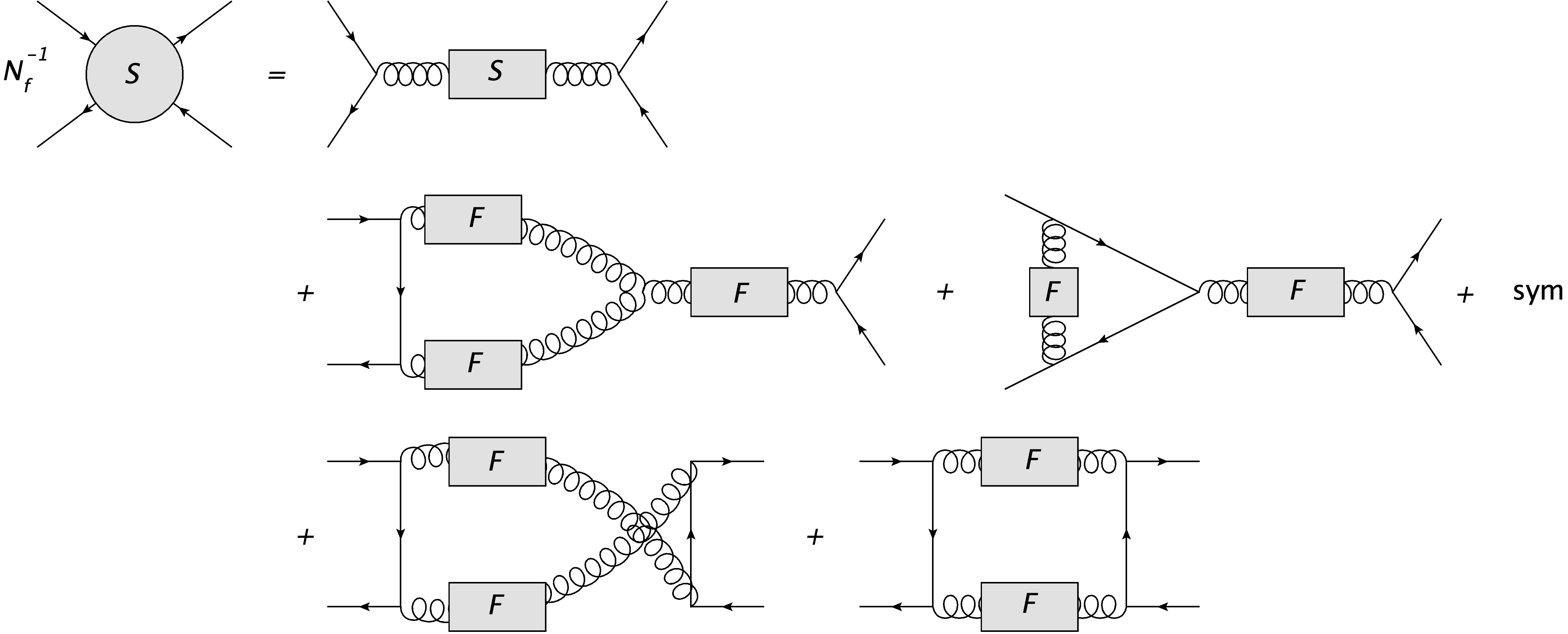}}
\noindent
where the triangle-based corrections in the second line should be symmetrized over the two quark pairs.
We denote by $N_f^{-1} K_{S,12}$ the bare subleading contribution arising from the first diagram above, and by $N_f^{-1} K_{S,3...6}$ the next four diagrams, respectively.
Similarly as for the gluon propagator, it does not make sense to add diagrams with a different loop counting, so let us analyse these contributions case by case.

\subsubsection{Tree-level-based}
The tree-level-based diagram receives contributions from all the subleading corrections to the full gluon propagator $D_S$ discussed in sec.~\eqref{sec:gg}

\centerline{\includegraphics[width=0.3\textwidth]{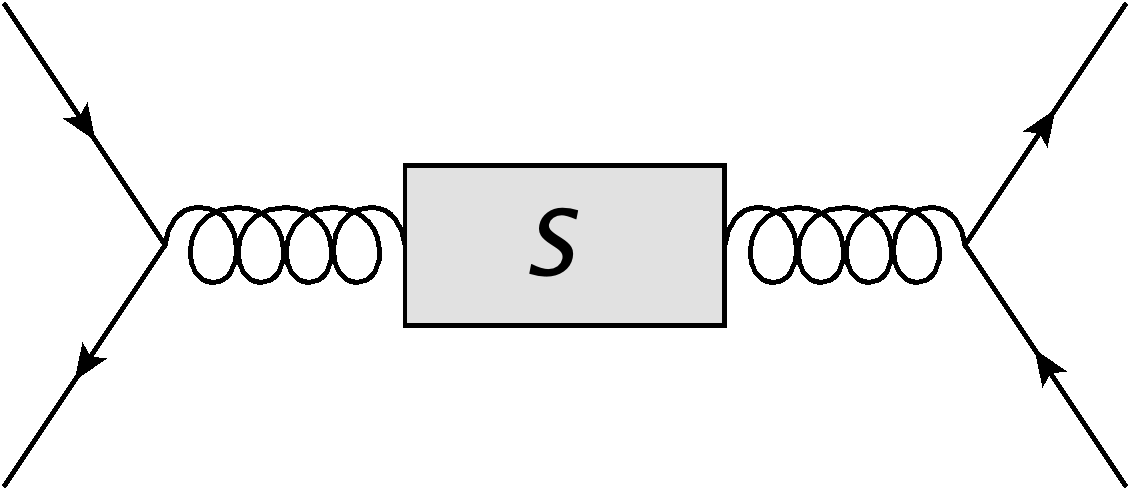}}
\begin{equation}
\begin{split}
= K_{S,12} 
= i g^2 \left( \frac{\mathcal{C}_1}{N_c} - \mathcal{C}_2 \right) 
\sum_{t,i=1}^{42} (A^{-1})_{ti} \, F_{L,i}(s,t,u,p_1^2,\ep) \, T_t \, D_S \,.
\end{split}
\end{equation}

\subsubsection{One-loop-triangle-based}

Consider the one-loop-triangle-based contributions to the all-loop BS kernel at subleading order in large $N_f$.

\paragraph{Gluon vertex correction diagram}\mbox{}\\
The gluon vertex correction diagram reads

\centerline{\includegraphics[width=0.4\textwidth]{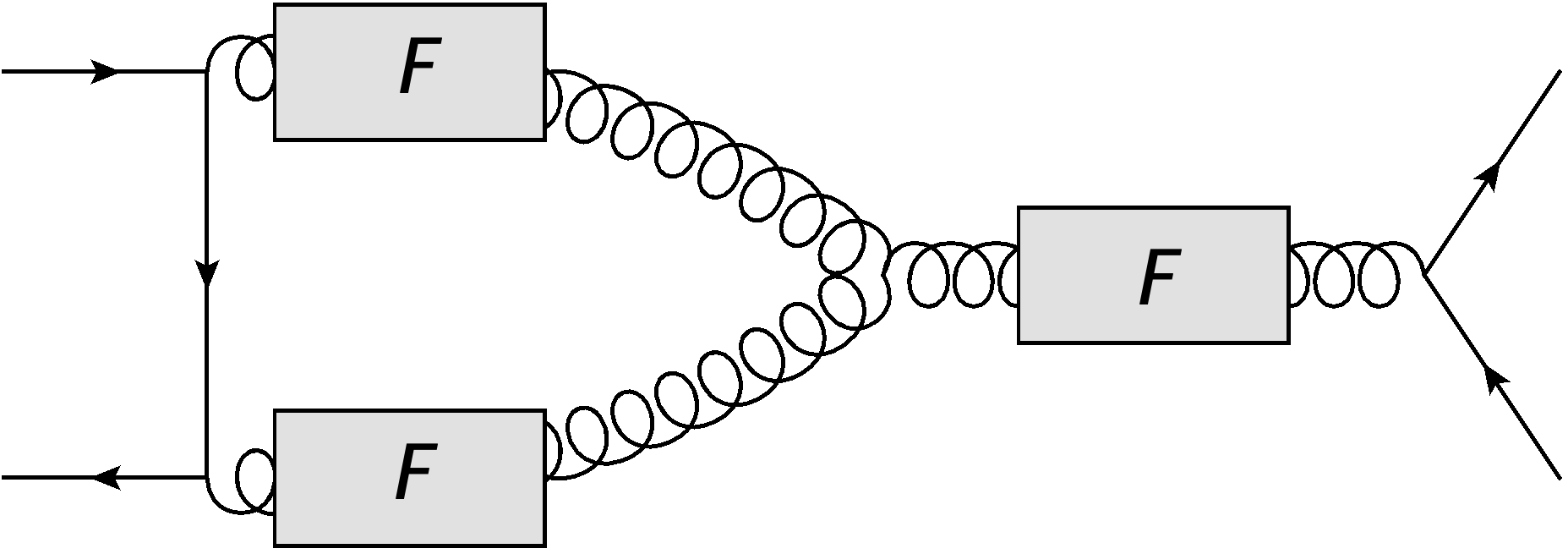}}
\begin{equation}
\begin{split}
&= N_f^{-1} K_{S,3} = \\
&= \frac{g^2 \lambda}{N_f} 
\left( \mathcal{C}_1 -N_c \, \mathcal{C}_2 \right)
\sum_{n,m=0}^\infty \int \frac{d^d k}{(2\pi)^d}
\frac{\mathcal{N}_{K,3}(k,p_1,p_2,p_3,\ep)}{(-k^2)(-(k+p_1)^2)(-(k+p_{12})^2)} \\
&\times \left( \frac{\Pi_{L}(k^2)}{k^2} \right)^n \left( \frac{\Pi_{L}((k+p_{12})^2)}{(k+p_{12})^2} \right)^m \sum_{l=0}^\infty \left( \frac{\Pi_{L}(s)}{s} \right)^l \\
&\overset{\text{IBP}}{=\joinrel=} \frac{g^2 \lambda}{N_f} 
\left( \mathcal{C}_1 -N_c \, \mathcal{C}_2 \right)  
\sum_{n,m,l=0}^\infty 
(\lambda \, b(\ep))^{n+m} \\
&\times \sum_{i=1}^2 c_{K,3,i}(s,t,u,p_1^2,\ep,n,m) M^{(K,3)}_i(s,t,u,p_1^2,\ep,n,m) \left( \frac{\Pi_{L}(s)}{s} \right)^l \,,
\end{split}
\end{equation}
where the numerator $\mathcal{N}$ involves all the color, Lorentz, and Dirac tensor structures.
We define the integral topology
\begin{equation}
\{ \mathcal{D}_i^{(\text{2mh)}} \} = \{ -k^2 ,  -(k+p_1)^2 ,  -(k+p_{12})^2 , -(k+p_{123})^2 \} \,.
\end{equation}
As the MI coefficients become much more involved rational functions than in the case of the gluon and quark propagators, we will not explicitly show their form here.
The MIs are
\begin{equation}
M_1^{(K,3)}(s)
= I_{1+n\ep,0,1+m\ep,0}^{(\text{2mh})}
= I_{1+n\ep,1+m\ep}^{(L)}|_{p^2 \to s}
= M^{(S,1,2)}(s)
\end{equation}

\centerline{\includegraphics[width=0.2\textwidth]{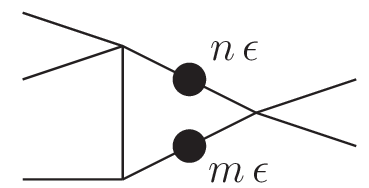}}
\begin{equation}
\begin{split}
&= M_2^{(K,3)}(s,p_1^2) 
= I_{1+n\ep,1,1+m\ep,0}^{(\text{2mh})}\\ 
&= \frac{i}{(4\pi)^{2-\ep}}
\frac{p_1^2 \Gamma (-m \epsilon -\epsilon ) \Gamma (-n \epsilon -\epsilon ) \Gamma (n \epsilon +\epsilon +1) s^{-m \epsilon -n \epsilon -\epsilon -2}  }{\Gamma (m \epsilon +1) \Gamma (n \epsilon +1) \Gamma (-m \epsilon -n \epsilon -2 \epsilon +1)} \left(\frac{p_1^2}{s}\right){}^{-n \epsilon -\epsilon -1} \\
&\times \left(\Gamma (m \epsilon +1)-\frac{\Gamma (m \epsilon +n \epsilon +\epsilon +1) B_{\frac{p_1^2}{s}}(n \epsilon +\epsilon ,m \epsilon
	+1)}{\Gamma (n \epsilon +\epsilon )}\right) 
    \left(1-\frac{p_1^2}{s}\right){}^{-m \epsilon -1} \\
&= \frac{i}{(4\pi)^{2}}
\frac{1}{(m+1) (n+1) \epsilon ^2 \left(s-p_1^2\right)}
+ \mathcal{O}(\ep^{-1}) \,,
\end{split}
\end{equation}
where $B_{\frac{p_1^2}{s}}(n \epsilon +\epsilon ,m \epsilon+1)$ is the incomplete Beta function.
The result is
\begin{equation}
\begin{split}
&N_f^{-1} K_{S,3} = \frac{g^2 \lambda}{N_f} 
\left( \mathcal{C}_1 -N_c \, \mathcal{C}_2 \right)  
\sum_{n,m,l=0}^\infty 
(\lambda \, b(\ep))^{n+m}
\left( \frac{\Pi_{L}(s)}{s} \right)^l \\
&\times \frac{i \, e^{-\ep \gamma_E}}{(4\pi)^{2-\ep}}
\sum_{t,i=1}^{42} (A^{-1})_{ti} \, F_{S,3,i}(s,t,u,p_1^2,\ep,n,m) \, T_t 
+ \mathcal{O}(\ep) \,,
\end{split}
\end{equation}
where $F_{S,3,i}$ are attached in the ancillary files.

\paragraph{Quark vertex correction diagram}\mbox{}\\
The quark vertex correction diagram reads

\centerline{\includegraphics[width=0.3\textwidth]{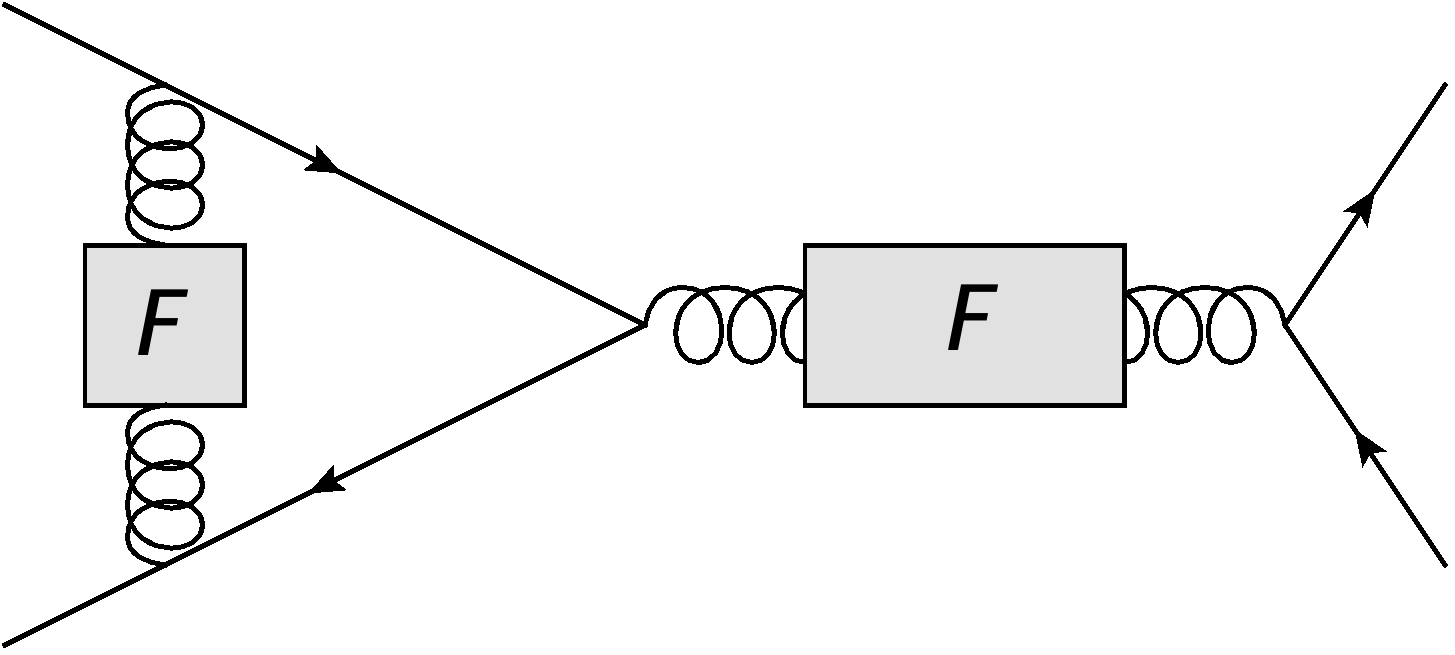}}
\begin{equation}
\begin{split}
&= N_f^{-1} K_{S,4} = \\
&= \frac{g^2 \lambda}{N_f}
\left( \frac{\mathcal{C}_1}{N_c^2} -\frac{\mathcal{C}_2}{N_c} \right) 
\sum_{n=0}^\infty \int \frac{d^d k}{(2\pi)^d}
\frac{\mathcal{N}_{K,4}(k,p_1,p_2,p_3,\ep)}{(-k^2)(-(k+p_1)^2)(-(k+p_{12})^2)} \\
&\times \left( \frac{\Pi_{L}((k+p_1)^2)}{(k+p_1)^2} \right)^n \sum_{m=0}^\infty \left( \frac{\Pi_{L}(s)}{s} \right)^m \\
&\overset{\text{IBP}}{=\joinrel=} \frac{g^2 \lambda}{N_f} 
\left( \frac{\mathcal{C}_1}{N_c^2} -\frac{\mathcal{C}_2}{N_c} \right) 
\sum_{n,m=0}^\infty 
(\lambda \, b(\ep))^{n} \\
&\times \sum_{i=1}^2 c_{K,4,i}(s,t,u,p_1^2,\ep,n) M^{(K,4)}_i(s,t,u,p_1^2,\ep,n) \left( \frac{\Pi_{L}(s)}{s} \right)^m \,,
\end{split}
\end{equation}
with MIs
\begin{equation}
M_1^{(K,4)}(p_1^2)
= I_{1,1+n\ep,0,0}^{(\text{2mh})}
= I_{1,1+n\ep}^{(L)}|_{p^2 \to p_1^2}
= M^{(S,1,2)}|_{m=0}(p_1^2)
\end{equation}

\centerline{\includegraphics[width=0.2\textwidth]{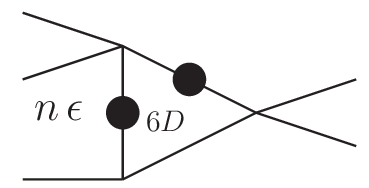}}
\begin{equation}
\begin{split}
&= M_2^{(K,4)}
= I_{2,1+n\ep,1,0}^{(\text{2mh,6D})} \\ 
&= \frac{i}{(4\pi)^{2}}
\frac{\log(-s)-\log\left(-p_1^2\right)}{s-p_1^2}
+ \mathcal{O}(\ep) \,.
\end{split}
\end{equation}
The result is
\begin{equation}
\begin{split}
&N_f^{-1} K_{S,4} = \frac{g^2 \lambda}{N_f}
\left( \frac{\mathcal{C}_1}{N_c^2} -\frac{\mathcal{C}_2}{N_c} \right) 
\sum_{n,m=0}^\infty 
(\lambda \, b(\ep))^{n}
\left( \frac{\Pi_{L}(s)}{s} \right)^m \\
&\times \frac{i \, e^{-\ep \gamma_E}}{(4\pi)^{2-\ep}}
\sum_{t,i=1}^{42} (A^{-1})_{ti} \, F_{S,4,i}(s,t,u,p_1^2,\ep,n) \, T_t 
+ \mathcal{O}(\ep) \,,
\end{split}
\end{equation}
where $F_{S,4,i}$ are attached in the ancillary files.

\subsubsection{One-loop-box-based}

Consider the one-loop-box-based contributions to the all-loop BS kernel at subleading order in large $N_f$.

\paragraph{Two-mass-easy diagram}\mbox{}\\
The two-mass-easy diagram reads

\centerline{\includegraphics[width=0.3\textwidth]{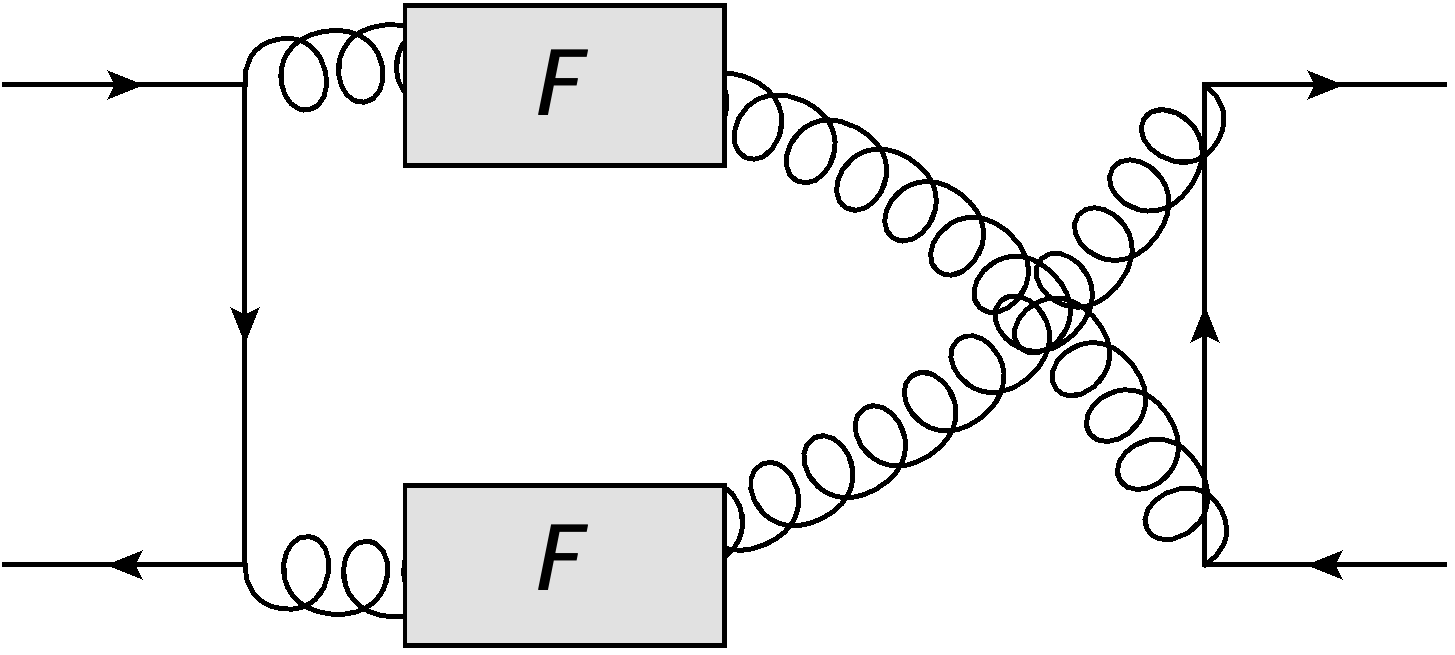}}
\begin{equation}
\begin{split}
&= N_f^{-1} K_{S,5} = \\
&= \frac{g^2 \lambda}{N_f} 
\left( \left( 1 + \frac{1}{N_c^2} \right) \mathcal{C}_1 -\frac{2}{N_c} \mathcal{C}_2 \right) \\
&\times \sum_{n,m=0}^\infty \int \frac{d^d k}{(2\pi)^d}
\frac{\mathcal{N}_{K,6}(k,p_1,p_2,p_3,\ep)}{(-k^2)(-(k+p_1)^2)(-(k+p_{12})^2) (-(k+p_{124})^2)} \\
&\times \left( \frac{\Pi_{L}(k^2)}{k^2} \right)^n \left( \frac{\Pi_{L}((k+p_{12})^2)}{(k+p_{12})^2} \right)^m \\
&\overset{\text{IBP}}{=\joinrel=} \frac{g^2 \lambda}{N_f} 
\left( \left( 1 + \frac{1}{N_c^2} \right) \mathcal{C}_1 -\frac{2}{N_c} \mathcal{C}_2 \right)
\sum_{n,m=0}^\infty 
(\lambda \, b(\ep))^{n+m} \\
&\times \sum_{i=1}^5 c_{K,5,i}(s,t,u,p_1^2,\ep,n,m)
M^{(K,5)}_i(s,t,u,p_1^2,\ep,n,m) \,.
\end{split}
\end{equation}
We define the integral topology
\begin{equation}
\{ \mathcal{D}_i^{(2me)} \} = \{ -k^2 ,  -(k+p_1)^2 ,  -(k+p_{12})^2 , -(k+p_{124})^2 \} \,.
\end{equation}
The MIs are
\begin{equation}
	M_1^{(K,5)}(s)
	= I_{1+n\ep,0,1+m\ep,0}^{(\text{2me})}
	= I_{1,1+n\ep}^{(L)}|_{p^2 \to s}
	= M^{(S,1,2)}(s)
\end{equation}

\begin{equation}
	M_2^{(K,5)}(s,p_4^2)
	= I_{1+n\ep,0,1+m\ep,1}^{(\text{2me})}
	= I_{1+n\ep,1,1+m\ep,0}^{(\text{2mh})}|_{p_1^2 \leftrightarrow p_4^2}
	= M^{(K,3)}_2(s,p_4^2)
\end{equation}

\begin{equation}
	M_3^{(K,5)}(s,p_1^2)
	= I_{1+n\ep,1,1+m\ep,0}^{(\text{2me})}
	= I_{1+n\ep,1,1+m\ep,0}^{(\text{2mh})}
	= M^{(K,3)}_2(s,p_1^2)
\end{equation}

\centerline{\includegraphics[width=0.2\textwidth]{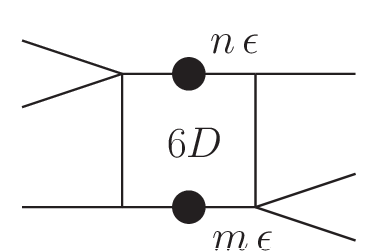}}
\begin{equation}
\begin{split}
&= M_4^{(K,5)}
= I_{1+n\ep,1,1+m\ep,1}^{(\text{2me,6D})} 
=  \mathcal{O}(\ep^0)
\end{split}
\end{equation}

\centerline{\includegraphics[width=0.2\textwidth]{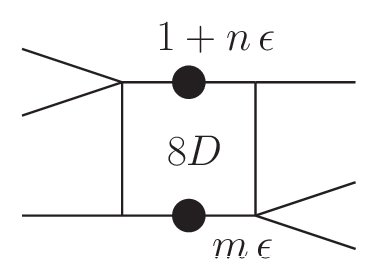}}
\begin{equation}
\begin{split}
&= M_5^{(K,5)}
= I_{2+n\ep,1,1+m\ep,1}^{(\text{2me,8D})}
=  \mathcal{O}(\ep^0) \,,
\end{split}
\end{equation}
where we do not explicitly provide here the expressions for the box MIs.
The result is
\begin{equation}
\begin{split}
&N_f^{-1} K_{S,5} = \frac{g^2 \lambda}{N_f}
\left( \left( 1 + \frac{1}{N_c^2} \right) \mathcal{C}_1 -\frac{2}{N_c} \mathcal{C}_2 \right)
\sum_{n,m=0}^\infty 
(\lambda \, b(\ep))^{n+m} \\
&\times \frac{i \, e^{-\ep \gamma_E}}{(4\pi)^{2-\ep}}
\sum_{t,i=1}^{42} (A^{-1})_{ti} \, F_{S,5,i}(s,t,u,p_1^2,\ep,n,m) \, T_t 
+ \mathcal{O}(\ep) \,,
\end{split}
\end{equation}
where $F_{S,5,i}$ are attached in the ancillary files.

\paragraph{Two-mass-hard diagram}\mbox{}\\
The two-mass-hard diagram reads

\centerline{\includegraphics[width=0.25\textwidth]{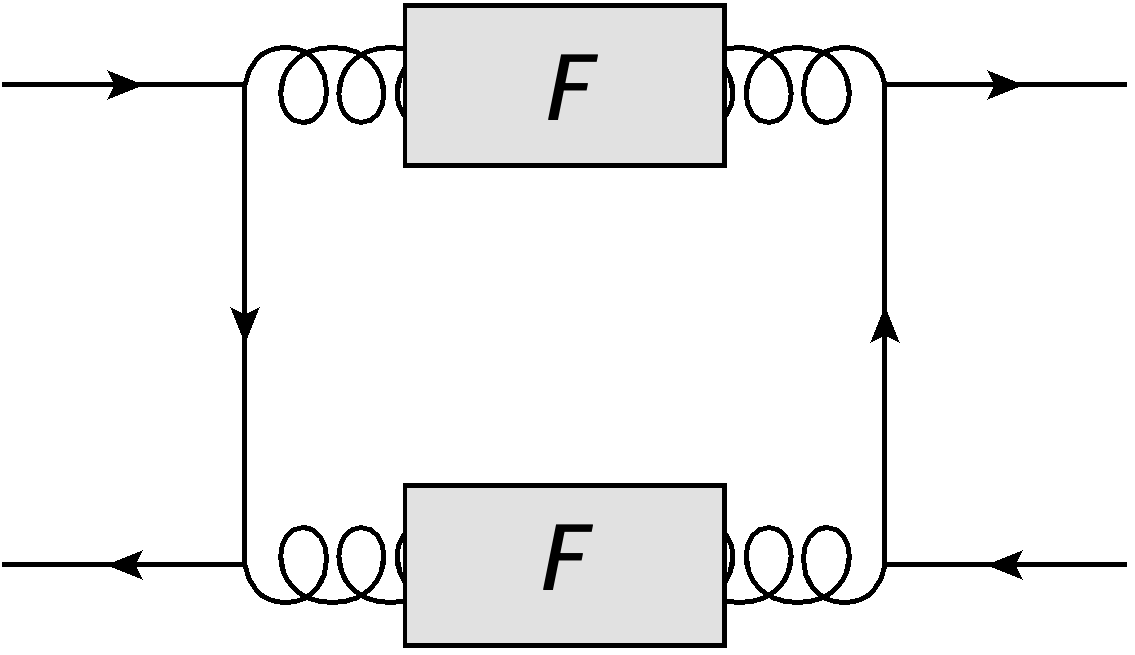}}
\begin{equation}
\begin{split}
&= N_f^{-1} K_{S,6} = \\
&= \frac{g^2 \lambda}{N_f} 
\left( \frac{1}{N_c} \mathcal{C}_1 + \left( N_c - \frac{2}{N_c} \right) \mathcal{C}_2 \right) \\
&\times \sum_{n,m=0}^\infty \int \frac{d^d k}{(2\pi)^d}
\frac{\mathcal{N}_{K,6}(k,p_1,p_2,p_3,\ep)}{(-k^2)(-(k+p_1)^2)(-(k+p_{12})^2) (-(k+p_{123})^2)} \\
&\times \left( \frac{\Pi_{L}(k^2)}{k^2} \right)^n \left( \frac{\Pi_{L}((k+p_{12})^2)}{(k+p_{12})^2} \right)^m \\
&\overset{\text{IBP}}{=\joinrel=} \frac{g^2 \lambda}{N_f} 
\left( \frac{1}{N_c} \mathcal{C}_1 + \left( N_c - \frac{2}{N_c} \right) \mathcal{C}_2 \right) \\
&\times \sum_{n,m=0}^\infty 
(\lambda \, b(\ep))^{n+m} 
\sum_{i=1}^6 c_{K,6,i}(s,t,u,p_1^2,\ep,n,m)
M^{(K,6)}_i(s,t,u,p_1^2,\ep,n,m) \,,
\end{split}
\end{equation}
with MIs
\begin{equation}
M_1^{(K,6)}(s)
= I_{1+n\ep,0,1+m\ep,0}^{(\text{2mh})}
= I_{1+n\ep,1+m\ep}^{(L)}|_{p^2 \to s}
= M^{(S,1,2)}(s)
\end{equation}

\begin{equation}
M_2^{(K,6)}(s,p_4^2)
= I_{1+n\ep,0,1+m\ep,1}^{(\text{2mh})}
= I_{1+n\ep,1,1+m\ep,0}^{(\text{2mh})}|_{p_1^2 \leftrightarrow p_4^2}
= M^{(K,3)}_2(s,p_4^2)
\end{equation}

\begin{equation}
M_3^{(K,6)}
= I_{1+n\ep,1,1+m\ep,0}^{(\text{2mh})}
= M^{(K,3)}_2
\end{equation}

\centerline{\includegraphics[width=0.2\textwidth]{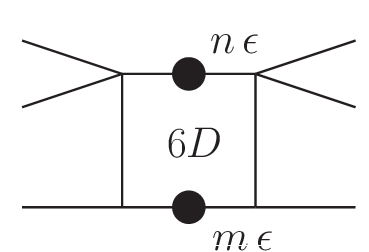}}
\begin{equation}
\begin{split}
&= M_4^{(K,6)}
= I_{1+n\ep,1,1+m\ep,1}^{(\text{2mh,6D})} 
= \mathcal{O}(\ep)
\end{split}
\end{equation}

\centerline{\includegraphics[width=0.2\textwidth]{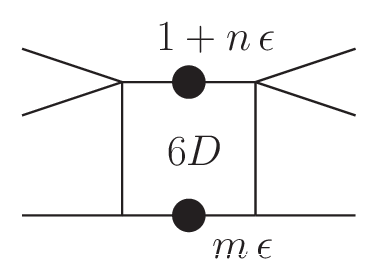}}
\begin{equation}
\begin{split}
&= M_5^{(K,6)}
= I_{2+n\ep,1,1+m\ep,1}^{(\text{2mh,6D})} 
= \mathcal{O}(\ep)
\end{split}
\end{equation}

\centerline{\includegraphics[width=0.2\textwidth]{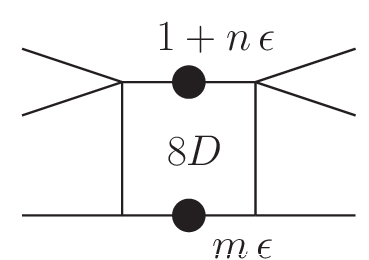}}
\begin{equation}
\begin{split}
&= M_6^{(K,6)}
= I_{2+n\ep,1,1+m\ep,1}^{(\text{2mh,8D})} 
= \mathcal{O}(\ep) \,.
\end{split}
\end{equation}
Note that the two-mass-hard box is linearly reducible in the following kinematic variables
\begin{equation}
s_{12} = x \, s_{14} \,, \qquad
p_4^2 = y \, \bar{y} \, s_{14} \,, \qquad
p_1^2 = (1-y) \, (1-\bar{y}) \, s_{14} \,.
\end{equation}
The result is
\begin{equation}
\begin{split}
&N_f^{-1} K_{S,6} = \frac{g^2 \lambda}{N_f}
\left( \frac{1}{N_c} \mathcal{C}_1 + \left( N_c - \frac{2}{N_c} \right) \mathcal{C}_2 \right)
\sum_{n,m=0}^\infty 
(\lambda \, b(\ep))^{n+m} \\
&\times \frac{i \, e^{-\ep \gamma_E}}{(4\pi)^{2-\ep}}
\sum_{t,i=1}^{42} (A^{-1})_{ti} \, F_{S,6,i}(s_{14},x,y,\bar{y},\ep,n,m) \, T_t 
+ \mathcal{O}(\ep) \,,
\end{split}
\end{equation}
where $F_{S,6,i}$ are attached in the ancillary files.

\subsubsection{Results}

The resulting one-loop-based functions $F_{S}$ have the following form
\begin{equation}
F_{S}(s,t,u,p_1^2,\ep,n,m) = \sum_{j=-2}^0 \ep^{j} \, \sum_{\alpha} a_{j,\alpha}(s,t,u,p_1^2,n,m) \ G_\alpha(s,t,u,p_1^2)
+ \mathcal{O}(\ep) \,.
\end{equation}
They are a series in $\ep$ involving both UV and IR poles $\ep^{-1}$, as well as IR poles $\ep^{-2}$~\footnote{Note that the IR poles would still be present if the quarks were treated as massive.}.
The coefficients $a_{j,\alpha}$ are rational functions of the kinematic invariants $\{s,t,u,p_1^2\}$ and the loop indices $\{n,m\}$.
The transcendental functions $G_{\alpha}$ are GPLs of kinematic invariants $\{s,t,u,p_1^2\}$.
\begin{figure}[h]
\centering
\begin{subfigure}[b]{0.49\textwidth}
\includegraphics[width=\textwidth]{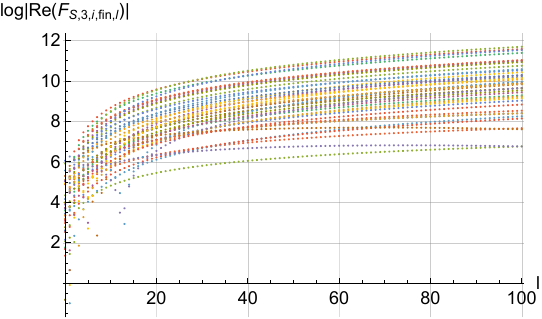}
\end{subfigure}
\begin{subfigure}[b]{0.49\textwidth}
\includegraphics[width=\textwidth]{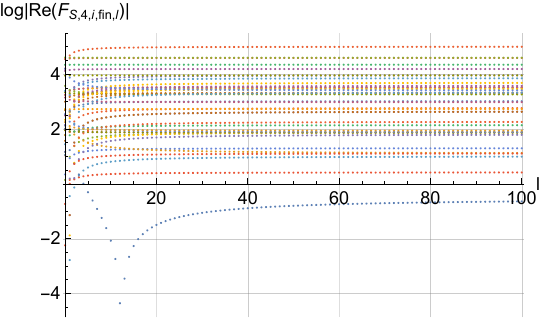}
\end{subfigure}
\begin{subfigure}[b]{0.49\textwidth}
\includegraphics[width=\textwidth]{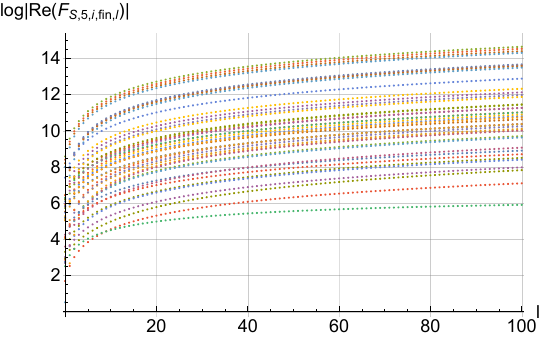}
\end{subfigure}
\begin{subfigure}[b]{0.49\textwidth}
\includegraphics[width=\textwidth]{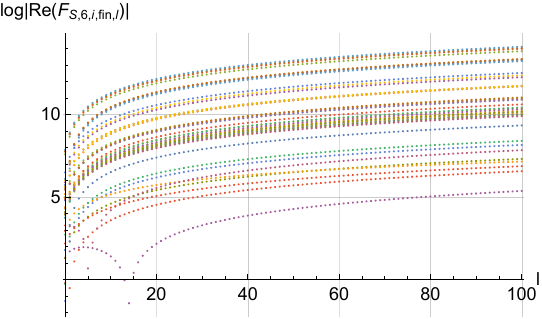}
\end{subfigure}
\caption{Finite part in $\ep$ of the $l$-loop bare massless subleading large $N_f$ BS kernel form factors $\log|F_{S,j,i,\text{fin},l}|$ as a function of $l$ for the four diagrams $j=3,4,5,6$ and all 42 tensors coefficients $i$ at kinematic point $s=-1 \,, u=-2.1 \,, t=-3.2 \,, p_1^2=-5.4 \,$.}
\label{fig:plots}
\end{figure}
Fig.~\eqref{fig:plots} presents the dependence of the one-loop-based form factors $F_{S}$ on the loop order $l$.
Note that $F_{S,4}$ has a logarithmic large $l$-loop asymptotic, while $\{F_{S,3} \,, F_{S,5} \,, F_{S,6}\}$ increase even faster with $l$.

\section{Conclusions and outlook}
\label{sec:concl}

In this work, we calculated the all-loop bare perturbative part of the gluon propagator, quark propagator, and four-quark BS kernel using modern scattering amplitude methods.

We perform the calculation to subleading order in the large number of quark flavors $N_f$ approximation of massless QCD.
This approximation simultaneously makes an all-loop calculation feasible, is systematically improvable, and preserves asymptotic freedom.
In order to perform our calculation analytically, we exploit state-of-the-art methods in Lorentz tensor decomposition, IBP reduction of Feynman integrals into a MI basis, and direct integration into GPLs.
As a byproduct of our BS kernel computation, we provide the gluon and quark propagator, also to subleading order in massless large $N_f$ QCD.
We provide our compact results explicitly in this manuscript, while the more complicated ones in a computer-readable format in the ancillary material that accompany this submission.
We also present the asymptotics of the all-loop summands at the large number of loops limit.

There are natural future directions following the discussion presented in this work.
First, one needs to extract an appropriate finite part of our bare all-loop perturbative results.
This can be achieved by a one-loop IR regularization~\cite{Catani:1998bh} of the BS kernel and an all-loop UV renormalization of both the kernel and propagators, following e.g. Refs~\cite{Palanques-Mestre:1983ogz,Gracey:1996he,Antipin:2018zdg}.
Second, the finite all-loop perturbative part of the correlator could be promoted to a fully nonperturbative result.
To this end, one could exploit transseries resurgence techniques~\cite{Dorigoni:2014hea,Dunne:2025mye}.
Third, the resulting nonperturbative correlator is an alternative first-principle solution to the DSEs~\cite{Radozycki:1995ax,Williams:2007zzh}.
Importantly, it also avoids the ambiguity of choosing a DS truncation scheme.
As such, it can be used as an input of the BSE for mesons~\cite{Fischer:2006ub,Swanson:2010pw}.
The mass spectrum resulting from the BSE solution could then be compared to experimentally measured values.
Fourth, one could improve the precision of our approximation by considering either quark masses or higher-order subleading corrections in large number of flavors $N_f$.
Finally, one could further extend this workflow to other observables.
For example, the baryonic spectrum~\cite{Eichmann:2016yit} would result from calculating a six-quark BS kernel, instead of our four-quark one.
From these mass spectra, one can extract the Regge trajectories~\cite{Hoyer:2016aew}.
In addition, there are BSE formulations that allow for computing e.g. the pion decay constant $f_\pi$~\cite{Pagels:1979hd,Eichmann:2016yit} and Parton Distribution Functions (PDFs)~\cite{Yu:2024ovn}.
We look forward to investigating these directions in the future.

\section*{Acknowledgements}

We are grateful to the Institute for Advanced Study for hospitality during the Prospects in Theoretical Physics 2023 graduate school, where the inspiration for this project was sparked.
We thank O. Antipin, G. Flacioni, C. Fischer, M. Hansen, F. Herren, F. Herzog, P. Hoyer, E. Gardi, P. Jakubčík, A. McLeod, E. Panzer, T. Radożycki, H. Sazdijan, K. Sch{\"o}nwald, Z. Zhu, and R. Zwicky for interesting discussions. 
The research of PB was supported by the Swiss National Science Foundation (SNF) under contract 200020-204200, by the European Research Council (ERC) under the European Union's Horizon 2020 research and innovation programme grant agreement 101019620 (ERC Advanced Grant TOPUP), and the ERC under the European Union’s Horizon Europe research and innovation program grant agreement 101163627 (ERC Starting Grant “AmpBoot”).
Feynman graphs were drawn with \texttt{Jaxodraw}~\cite{Binosi:2003yf,Vermaseren:1994je}.

\bibliographystyle{JHEP}
\bibliography{references}

\end{document}